\documentclass[reqno,a4paper,11pt]{article}
\pdfoutput=1
\usepackage{xcolor}

\usepackage{graphicx}
\usepackage[textwidth = 430 pt, textheight = 630 pt]{geometry}

\definecolor{MyDarkBlue}{rgb}{0.15,0.25,0.45}
\usepackage{epsfig,rotating}
\usepackage{amsmath,amssymb}
\usepackage{amsfonts}
\usepackage{mathrsfs}
\usepackage{bbm}
\usepackage[normalem]{ulem}

\usepackage{array,tabularx}

\usepackage{latexsym}
\usepackage{amsthm}

\usepackage[utf8x]{inputenc}

\usepackage{hyperref}
\hypersetup{
hypertexnames=false,
colorlinks=true,
citecolor=MyDarkBlue,
linkcolor=MyDarkBlue,
urlcolor=MyDarkBlue,
pdfauthor={Christian S\"amann},
pdftitle={Lectures on Higher Structures in M-theory},
pdfsubject={hep-th math-ph}
breaklinks=true
}

\usepackage{tikz}
\usepackage{mathtools}

\usepackage[all,knot]{xy}
\xyoption{arc}

\let\fn\footnote
\renewcommand{\footnote}[1]{\linespread{1.1}\fn{#1}\linespread{1.29}}

\makeatletter\renewcommand{\section}{\@startsection
{section}{1}{\z@}{-3.5ex plus -1ex minus
    -.2ex}{2.3ex plus .2ex}{\bf }}
\makeatletter\renewcommand{\subsection}{\@startsection{subsection}{2}{\z@}{-3.25ex
plus -1ex minus
   -.2ex}{1.5ex plus .2ex}{\bf }}
\makeatletter\renewcommand{\subsubsection}{\@startsection{subsubsection}{3}{-2.45ex}{-3.25ex
plus -1ex minus -.2ex}{1.5ex plus .2ex}{\it }}
\renewcommand{\thesection}{\arabic{section}}
\renewcommand{\thesubsection}{\arabic{section}.\arabic{subsection}}
\renewcommand{\@seccntformat}[1]{\@nameuse{the#1}.~~}

\renewcommand{\theequation}{\thesection.\arabic{equation}}
\makeatletter \@addtoreset{equation}{section}

\usepackage[toc,page]{appendix}

\newtheorem{thm}{Theorem}[section]
\renewcommand{\thethm}{\thesection.\arabic{thm}}

\newtheorem{theorem}[thm]{Theorem}

\newtheorem{remark}[thm]{Remark}

\renewcommand{\appendices}{
\section*{Appendix}\label{appendices}\setcounter{subsection}{0}
\addcontentsline{toc}{section}{Appendix}
\setcounter{equation}{0}
\makeatletter
\renewcommand{\theequation}{\Alph{subsection}.\arabic{equation}}
\renewcommand{\thesubsection}{\Alph{subsection}}
\renewcommand{\thethm}{\Alph{subsection}.\arabic{thm}}
\@addtoreset{equation}{subsection}
\@addtoreset{thm}{subsection}
\makeatother
}

\makeatletter\renewcommand{\subsection}{\@startsection{subsection}{2}{\z@}{-3.25ex
plus -1ex minus
   -.2ex}{1.5ex plus .2ex}{\it }}

\def\slasha#1{\setbox0=\hbox{$#1$}#1\hskip-\wd0\hbox to\wd0{\hss\sl/\/\hss}}

\def\periodb#1{\setbox0=\hbox{$#1$}#1\hskip-\wd0\hbox to\wd0{-}}
\newcommand{\myxymatrix}[1]{\vcenter{\vbox{\xymatrix{#1}}}}
\newcommand{\nablas}{\slasha{\nabla}}

\newcommand{\lbr}{(\hspace{-0.1cm}(}
\newcommand{\rbr}{)\hspace{-0.1cm})}

\newcommand{\unit}{\mathbbm{1}}   			% identity map/matrix
\newcommand{\id}{\mathrm{id}}   			% identity map/matrix
\newcommand{\CA}{\mathcal{A}}    			% cal-letters

\newcommand{\xd}{\dot{x}}

\newcommand{\CC}{\mathcal{C}}
\newcommand{\CCC}{\mathscr{C}}
\newcommand{\CCL}{\mathscr{L}}

\newcommand{\CCD}{\mathscr{D}}

\newcommand{\CF}{\mathcal{F}}

\newcommand{\CCG}{\mathscr{G}}
\newcommand{\CH}{\mathcal{H}}

\newcommand{\CK}{\mathcal{K}}

\newcommand{\CL}{\mathcal{L}}
\newcommand{\CM}{\mathcal{M}}

\newcommand{\CN}{\mathcal{N}}
\newcommand{\CO}{\mathcal{O}}

\newcommand{\CCP}{\mathscr{P}}

\newcommand{\CT}{\mathcal{T}}

\newcommand{\frg}{\mathfrak{g}}				% frak-letters
\newcommand{\frh}{\mathfrak{h}}				% frak-letters

\newcommand{\FR}{\mathbbm{R}}     			% field of real numbers
\newcommand{\FC}{\mathbbm{C}}     			% field of complex numbers
\newcommand{\NN}{\mathbbm{N}}     			% set of natural numbers
\newcommand{\RZ}{\mathbbm{Z}}     			% ring of integers
\newcommand{\CPP}{{\mathbbm{C}P}}    			% complex projective plane
\newcommand{\dd}{\mathrm{d}}     			% total differential
\newcommand{\dpar}{\partial}     			% partial differential
\newcommand{\embd}{{\hookrightarrow}}     		% embedded
\newcommand{\di}{\mathrm{i}}     			% imaginary unit
\newcommand{\eps}{{\varepsilon}}			% antisymmetric tensors
\newcommand{\sB}{\mathsf{B}}

\newcommand{\ald}{{\dot{\alpha}}}     			% dotted letters
\newcommand{\bed}{{\dot{\beta}}}

\newcommand{\eand}{{\qquad\mbox{and}\qquad}}     		% and etc. in equations
\newcommand{\ewith}{{\qquad\mbox{with}\qquad}}
\newcommand{\efor}{{\qquad\mbox{for}\qquad}}

\newcommand{\der}[1]{\frac{\dpar}{\dpar #1}}   		% partielle ableitung, 1 argument
\newcommand{\dder}[1]{\frac{\dd}{\dd #1}}   		% partielle ableitung, 1 argument
\newcommand{\tr}{\,\mathrm{tr}\,}     			% trace
\newcommand{\agl}{\mathfrak{gl}}     			% algebras

\newcommand{\au}{\mathfrak{u}}
\newcommand{\asu}{\mathfrak{su}}

\newcommand{\sU}{\mathsf{U}}     			% groups

\newcommand{\sG}{\mathsf{G}}

\newcommand{\sL}{\mathsf{L}}
\newcommand{\sHom}{\mathsf{Hom}}
\newcommand{\sLie}{\mathsf{Lie}}

\newcommand{\sH}{\mathsf{H}}

\newcommand{\sSpin}{\mathsf{Spin}}

\newcommand{\acton}{\vartriangleright}     			% span
\renewcommand{\remark}[1]{}     				% remark
\def\tyng(#1){\hbox{\tiny$\yng(#1)$}}			% small Young diagram
\def\tyoung(#1){\hbox{\tiny$\young(#1)$}}			% small Young diagram
\newcommand{\beq}{\begin{eqnarray}}
\newcommand{\eeq}{\end{eqnarray}}

\newcommand{\sfs}{{\sf s}}
\newcommand{\sft}{{\sf t}}

\newcommand{\CatCat}{\mathsf{Cat}}
\newcommand{\CatSet}{\mathsf{Set}}

\newenvironment{exercise}{\noindent\rule[0.5ex]{\linewidth}{0.5pt} \wpem{Exercise:} }{\\ \noindent\rule[0.5ex]{\linewidth}{0.5pt}}

\newcommand{\wpcite}[1]{~\cite{#1}}
\newcommand{\wpeg}{e.g.}
\newcommand{\wpie}{i.e.}
\newcommand{\wpcf}{cf.}
\newcommand{\wpetal}{et al.}
\newcommand{\wpem}[1]{{\em #1}\new@ifnextchar\@sptoken{\/}{}}

\begin{document}
\begin{titlepage}
\begin{flushright}
 EMPG--16--17
\end{flushright}
\vskip 2.0cm
\begin{center}
{\LARGE \bf Lectures on Higher Structures in M-Theory}
\vskip 1.5cm
{\Large Christian S\"amann}
\setcounter{footnote}{0}
\renewcommand{\thefootnote}{\arabic{thefootnote}}
\vskip 1cm
\wpem{Maxwell Institute for Mathematical Sciences\\
Department of Mathematics, Heriot-Watt University\\
Colin Maclaurin Building, Riccarton, Edinburgh EH14 4AS, U.K.}\\[0.5cm]
{Email: {\ttfamily c.saemann@hw.ac.uk}}
\end{center}
\vskip 1.0cm
\begin{center}
{\bf Abstract}
\end{center}
\begin{quote}
These are notes for four lectures on higher structures in M-theory as presented at workshops at the Erwin Schrödinger Institute and Tohoku University. The first lecture gives an overview of systems of multiple M5-branes and introduces the relevant mathematical structures underlying a local description of higher gauge theory. In the second lecture, we develop the corresponding global picture. A construction of non-abelian superconformal gauge theories in six dimensions using twistor spaces is discussed in the third lecture. The last lecture deals with the problem of higher quantization and its relation to loop space. An appendix summarizes the relation between 3-Lie algebras and Lie 2-algebras.
\end{quote}
\end{titlepage}

\tableofcontents

\newpage

\section{Introduction}

%\todo{adjust journal/arXiv versions!}

The purpose of these lectures is to provide a motivation for studying higher structures within M-theory and to give some feeling for the mathematical language underlying these structures. In particular, we will review a very general approach to principal bundles with connections, which allows for a very large class of generalizations, including higher gauged sigma models on presentable higher differential stacks. This includes higher gauge theories on ordinary manifolds, orbifolds and categorified spaces. Our approach yields the global geometric bundle structure, the appropriate definition of curvatures and the finite gauge transformations.

My personal interest in such constructions stems from the problem of finding some description of the superconformal field theory in six dimensions with non-abelian gauge structure. This theory is usually referred to as the $(2,0)$-theory, as it has $\CN=(2,0)$ supersymmetry. It is well known that the free or abelian $(2,0)$-theory contains a 2-form curvature on a $\sU(1)$-gerbe. Therefore it is natural to expect that a non-abelian theory is based on non-abelian gerbes, which are categorifications of principal 2-bundles. In order to look for a non-abelian $(2,0)$-theory, one should first understand the geometric setup for the gauge sector.

Once this is done, there is a very natural and direct route to constructing corresponding $(2,0)$-theories. Recall that solutions to the $\CN=4$ Yang--Mills equations in four dimensions can be described by certain holomorphic principal bundles over a suitable twistor space. Similarly, a twistor space for self-dual 3-form curvatures in six dimensions is known. It is easy to extend this picture supersymmetrically, and we can consider more general holomorphic higher principal bundle over the resulting twistor space. In the case of holomorphic principal 2-bundles, this yields precisely the field content of the $(2,0)$-theory, together with superconformal field equations.

In a second part, I discuss higher quantization, a topic which is closely related to the definition of the $(2,0)$-theory. Essentially, one would expect that the higher endomorphisms on the resulting higher Hilbert spaces form the higher gauge algebras for the $(2,0)$-theory, as well as for generalized M2-brane models. 

Even though we focus on higher gauge theory and quantization in these notes, the mathematical tools presented are applicable in many contexts in string theory. One currently very active area of research that certainly could benefit from categorified geometric structures is double field theory.

Since I had only four hours available for the above material, the discussion is necessarily very concise. I tried give impressions of the right point of view of defining various mathematical objects, which is sometimes also called the $n$-categorical point of view or \href{http://ncatlab.org/nlab/show/nPOV}{$n$-POV}. This $n$-POV is the one unifying mathematical objects and their categorification (as well as physics, philosophy, you name it). 

The time constraint meant that some of the discussion had to remain mathematically superficial or even slightly sloppy. For example, I always avoided talking about internalization, which underlies the transition from $n$-groups to Lie $n$-groups. A complete treatment of all the issues is found in the references or on the \href{http://ncatlab.org}{nlab}. The latter webpage is also an invaluable source for mathematical definitions related to higher gauge theory and their applications in physics.

I have resisted the temptation to add material which I did not have time to discuss in the lectures. I took, however, the liberty of adding footnotes, literature references and remarks, providing additional connections between the discussed topics and objects. Also, an appendix addresses the frequently arising question of how the 3-Lie algebras of M2-brane models are related to the categorified Lie algebras used in our constructions.

\section*{Acknowledgements}

I would like to thank the organizers of the workshops ``Higher Structures in String Theory and Quantum Field Theory'' at the Erwin Schrödinger International Institute for Mathematical Physics, Vienna, as well as ``Higher Structures in String Theory and M-Theory'' at Tohoku University, Sendai, during which these lectures were presented. Many thanks also to Patricia Ritter, Lennart Schmidt and Martin Wolf for helpful comments on a first version of these notes.

\section{Higher gauge theory arising in M-theory}\label{sec:2}

In this section, we present our motivation for developing higher gauge theory. We introduce higher Lie algebras and give a glimpse of higher gauge theory by presenting its local description.

\subsection{Motivation: Systems of multiple M5-branes}

In string theory, interesting vacua are described in terms of D-brane configurations in certain background geometries with fluxes. The D-branes interact through strings stretched between them. The endpoints of these strings induce a $\sU(1)$-gauge field on the D-branes' worldvolumes and if we decouple massive modes and gravity, we obtain an effective dynamical description in terms of gauge theory. For a very helpful review of the arising gauge theories, see Giveon \& Kutasov\wpcite{Giveon:1998sr}.

When multiple D-branes come together, the abelian gauge symmetry is generally enhanced to a non-abelian one. In the simplest case of a flat stack of $n$ D-branes, the gauge group $\sU(1)\times \ldots \times \sU(1)$ is enhanced to $\sU(n)$. Non-abelian gauge theories are certainly much richer than abelian ones, because they exhibit new features as \wpeg\ confinement. But even without matter coupling, we have instantons in four dimensions and non-singular monopoles on $\FR^3$, which do not exist in the abelian case.

A particularly interesting example of such a gauge theory is $\CN=4$ super Yang--Mills theory on Minkowski space $\FR^{1,3}$. This theory arises from a stack of flat D3-branes and correspondingly has 16 real supersymmetries. Its classical superconformal symmetry is preserved at the quantum level, and it is the conformal field theory in the famous holographic correspondence on ${\rm AdS}_5\times S^5$. Because of its simplicity and accessibility, it is also dubbed the ``harmonic oscillator of the 21st century.''

In M-theory, the situation is similar, see \wpeg\ the review of Berman\wpcite{Berman:2007bv} for a detailed account. We can describe interesting vacua as M-brane configurations in background geometries with fluxes. Contrary to the rich spectrum of D-branes in type II superstring theory, there are essentially only M2- and M5-branes in M-theory. M5-branes interact through M2-branes ending on them and their boundaries form tensionless ``self-dual'' strings. Because they are essentially massless, supergravity decouples from the dynamics of these strings. Effectively they should be described by a six dimensional $\CN=(2,0)$-supersymmetric conformal field theory\wpcite{Witten:1995zh,Strominger:1995ac,Witten:1995em}. This theory, which we well call $(2,0)$-theory for short in the following, is in some sense the six dimensional analogue of $\CN=4$ super Yang--Mills theory in four dimensions. 
The field content of the $(2,0)$-theory consists of the $\CN=(2,0)$-tensor multiplet in six dimensions. Its bosonic sector is given by a self-dual 3-form curvature $H:=\dd B=*H$ of a 2-form potential $B$ and five scalars, which are the Goldstone modes of the breaking of the symmetries from $\FR^{1,10}$ to $\FR^{1,5}\times \FR^5$ by the presences of the M5-brane. 

Mathematically, the 2-form potential $B$ is known to be the curving of an abelian gerbe or an abelian principal 2-bundle\wpcite{Gawedzki:1987ak,Freed:1999vc} and dynamical theories of such higher form potentials on higher principal bundles are known as higher gauge theories. An important problem is now the non-abelian generalization of these structures, which should underlie an effective description of stacks of multiple M5-branes. There is huge interest in such a theory in both the physics and the mathematics communities, as it would boost our understanding of M-theory and shed light on many issues, as \wpeg\ Alday, Gaiotto, and Tachikawa (AGT) relations, the Geometric Langlands duality and M-theory in general. The larger part of these lectures will deal with an approach towards defining a classical version of the (2,0)-theory.

\subsection{Obstacles to constructing classical non-abelian (2,0)-theories}

A point particle charged under a gauge group $\sG$ is described by its position and a point in a representation space of $\sG$. More precisely, it is described by a point in a corresponding associated vector bundles to a principal $\sG$-bundle. Recall that given a cover $\sqcup_i U_i\rightarrow M$ of a manifold $M$ in terms of local patches $U_i$, such a principal bundle is given by a set of transition functions $g_{ij}:U_{ij}\rightarrow \sG$ satisfying 
\begin{equation}\label{eq:1-cocycle}
 g_{ij}g_{jk}=g_{ik}
\end{equation}
on all non-empty triple intersections $U_i\cap U_j\cap U_k$. Mathematically, these functions form a (non-abelian) \v Cech 1-cocycle. \v Cech 1-cocycles are extended to Deligne 1-cocycles by adding a set of $\sLie(\sG)$-valued connection 1-forms $A_i\in \Omega(U_i)\otimes\sLie(\sG)$, which are glued together by the relation
\begin{equation}
 A_j=g_{ij}^{-1}A_ig_{ij}+g_{ij}^{-1}\dd g_{ij}
\end{equation}
on non-empty double intersections $U_i\cap U_j$. Gauge transformations are parameterized by functions $\gamma_i:U_i\rightarrow \sG$, which form Deligne coboundaries between Deligne 1-cocycles $(g,A)$ and $(\tilde g,\tilde A)$ according to
\begin{equation}\label{eq:1-coboundary}
 \gamma_i\tilde g_{ij}=g_{ij}\gamma_j\eand \tilde A_i=\gamma_i^{-1}A_i\gamma_i+\gamma_i^{-1}\dd \gamma_i~.
\end{equation}
Deligne 1-cocycles modulo Deligne 1-coboundaries yield Deligne cohomology classes which describe principal $\sG$-bundles with connection modulo gauge transformations.

The potential one-forms $A_i$ give rise to a map from a path $\gamma$ in $M$ to a group element
\begin{equation}\label{eq:holonomy1}
 g(\gamma)=P\exp\int_\gamma A~,
\end{equation}
where $P$ stands for path ordering. This map encodes the parallel transport of $\sG$-charged particles along the path $\gamma$.

We can regard the curve $\gamma$ as the boundary of an open string on a D-brane, \wpie\ an endpoint of a string moving through time. After a lift to M-theory, we should consider analogously the boundary of an M2-brane ending on an M5-brane, which gives rise to the previously mentioned ``self-dual strings.'' To describe a parallel transport of self-dual strings along a surface $\sigma$, we are naturally led to introducing a 2-form $B$. In the abelian case, we can then write
\begin{equation}\label{eq:holonomy2}
 g(\sigma)=\exp\int_\sigma B~,
\end{equation}
but since there is no reparametrization invariant notion of surface ordering, this equation does not extend to the non-abelian case. More explicitly, when considering a parallel transport of a self-dual string subdivided into two pieces as follows:
\begin{subequations}\label{diag:interchange_law}
\begin{equation}
 \myxymatrix{ \bullet
&& 
  \ar@/_4ex/[ll]_{}="g1"
  \ar[ll]_(0.65){}
  \ar@{}[ll]|{}="g2"
  \ar@/^4ex/[ll]^{}="g3"
  \ar@{=>}^{g_1} "g1"+<0ex,-0.5ex>;"g2"+<0ex,0.5ex>
  \ar@{=>}^{g'_1} "g2"+<0ex,-0.5ex>;"g3"+<0ex,0.5ex>
&& \bullet
  \ar@/_4ex/[ll]_{}="h1"
  \ar[ll]_(0.65){}
  \ar@{}[ll]|{}="h2"
  \ar@/^4ex/[ll]^{}="h3"
  \ar@{=>}^{g_2} "h1"+<0ex,-0.5ex>;"h2"+<0ex,0.5ex>
  \ar@{=>}^{g'_2} "h2"+<0ex,-0.5ex>;"h3"+<0ex,0.5ex>
}
\end{equation}
we need that
\begin{equation}\label{eq:pre_interchange_law}
 (g'_1g'_2)(g_1g_2)=(g'_1g_1)(g'_2g_2)
\end{equation}
\end{subequations}
since the order in which we parallel transport the top string to the bottom one should be irrelevant. By an argument going back to Eckmann and Hilton\wpcite{Hilton:1962:227-255}, this forces $\sG$ to be abelian. It is essentially this argument that underlies many of the no-go-theorems in the physics literature which forbid the existence of a non-abelian parallel transport along surfaces.

Fortunately, the assumptions here are a little too naive. Note that the map \eqref{eq:holonomy1} from curves $\gamma$ to group elements is in fact part of a functor, \wpcf\ Mackaay \& Picken\wpcite{Mackaay:2000ac}, and it is very natural to assume that a corresponding map from surfaces should be a 2-functor. Indeed, the objects appearing in \eqref{diag:interchange_law} suggest that we have objects ($\bullet$), 1-morphisms ($\longleftarrow$) and 2-morphisms ($\Longleftarrow$). It is also clear from the diagram that the 1-morphisms can be composed horizontally, which induces a horizontal composition of the 2-morphisms. Both combine into a functor, which we will denote by $\otimes$. Moreover, there is a vertical composition $\circ$ of 2-morphisms. This is the structure appearing in the definition of a 2-category, and a direct consequence of the axioms for 2-categories is the appropriate form of Equation~\eqref{eq:pre_interchange_law},
\begin{equation}\label{eq:interchange_law}
 (g'_1\otimes g'_2)\circ (g_1\otimes g_2)=(g'_1\circ g_1)\otimes (g'_2\circ g_2)~,
\end{equation}
which is known as the \wpem{interchange law}.

Having removed the mathematical obstacles to constructing a classical $(2,0)$-theory, let us consider some arguments from string theory. Since the $(2,0)$-theory  is conformal we know that there are no dimensionful parameters in the theory. Contrary to $\CN=4$ super Yang--Mills theory, however, string theory considerations imply that the relevant superconformal fixed points in parameter space are isolated and therefore there are no continuous parameters either. This suggests that there is no Lagrangian description. 

Note, however, that the same arguments were true for M2-branes, and successful M2-brane models have been constructed\wpcite{Bagger:2007jr,Gustavsson:2007vu,Aharony:2008ug}. While there are no continuous parameters in these models, there is a discrete one, $k\in\RZ$, arising from the background geometry $\FR^8/\RZ_k$ in which the M2-branes are placed. We can expect that the same happens in the case of M5-branes.

Even if we were overly skeptical about the existence of a classical description of the $(2,0)$-theory, we would still expect a classical description of the BPS subsector of the theory to exist. Finally, even if this turned out to be false, we would still be able to learn interesting facts about the $(2,0)$-theory by studying quantum features of the non-abelian higher gauge theories that we will develop in the following.

\subsection{Self-dual strings}

Let us now try to gain some more intuition about the degrees of freedom underlying the $(2,0)$-theory. Consider the description of monopoles in type IIA superstring theory as D2-branes ending on D4-branes. The D2-brane and D4-brane are positioned in flat, ten-dimensional Minkowski space $\FR^{1,9}$ such that their worldvolumes fill the following directions:
\begin{equation}\label{diag:D2D4}
\begin{tabular}{rcccccccc}
& 0 & 1 & 2 & 3 & 4 & 5 & 6 & \ldots\\
D2 & $\times$ & & & & & $\times$ & $*$ \\
D4 & $\times$ & $\times$ & $\times$ & $\times$ & & $\times$ &
\end{tabular}
\end{equation}
Here the $*$ indicates that the D2-branes do not fill the entire $x^6$ direction but may end on the D4-branes. We are now interested in configurations which are constant in the time direction $x^0$ and the spatial direction $x^5$. From the perspective of the D4-brane, such configurations are described as follows. The effective D4-brane worldvolume theory is simply super Yang--Mills theory, and the presence of the D2-brane restricts the theory further to its BPS subsector, whose bosonic part is captured by the \wpem{Bogomolny monopole equation} on $\FR^3$,
\begin{equation}\label{eq:Bogomolny}
 F_{ij}=\eps_{ijk}\nabla_k\Phi~,~~~i,j,k=1,2,3~.
\end{equation}
Here, the field content for $n$ D4-branes is a gauge potential describing a connection $\nabla$ on a trivial principal $\sU(n)$-bundle over $\FR^3$ with curvature $F$ and the scalar field $\Phi$, taking values in the adjoint representation of $\sLie(\sU(n))$ describes the position of the D4-brane in the $x^6$-direction.

From the perspective of the D2-brane, we have an analogous description in terms of the BPS subsector of a Yang--Mills theory, described by the \wpem{Nahm equation}
\begin{equation}\label{eq:Nahm}
 \nabla_6 X^i=\tfrac{1}{2}\eps^{ijk}[X^j,X^k]~,
\end{equation}
where $\nabla_6=\der{x^6}+A_6$ and the gauge potential $A_6$ can be gauged away. Here, the three scalar fields $X^i$ encode the position of the D2-brane in the $x^{1,2,3}$-directions.

Interestingly, there is a duality between these two descriptions known as the Nahm transform. This includes the ADHMN-construction, which maps solutions to the Nahm equation to solutions to the Bogomolny monopole equation. This can be used to describe and study the moduli space of monopoles in a very efficient manner.

We can now lift the configuration~\eqref{diag:D2D4} up to M-theory, using the $x^4$-directions as the M-theory direction:
\begin{equation}\label{diag:M2M5}
\begin{tabular}{rccccccc}
${\rm M}$ & 0 & 1 & 2 & 3 & \phantom{(}4\phantom{)} & 5 & 6 \\
M2 & $\times$ & & & & & $\times$ & $*$ \\
M5 & $\times$ & $\times$ & $\times$ & $\times$ & $\times$ & $\times$ 
\end{tabular}
\end{equation}
In the abelian case of a single M2-brane, this configuration is described by the self-dual string equation\wpcite{Howe:1997ue}
\begin{equation}\label{eq:self-dual-string}
 H_{\mu\nu\kappa}:=\dpar_{[\mu}B_{\nu\kappa]}=\eps_{\mu\nu\kappa\lambda}\dpar_\lambda\Phi~,~~~\mu,\nu,\kappa,\lambda=1,\ldots,4~,
\end{equation}
where $B_{\mu\nu}$ is a potential 2-form on a trivial $\sU(1)$-gerbe over $\FR^4$. 

\begin{exercise}
 Show that a Kaluza--Klein reduction of Equation~\eqref{eq:self-dual-string} yields a gauge potential on $\FR^3$ satisfying Equation~\eqref{eq:Bogomolny} together with a gauge-trivial 2-form potential.
\end{exercise}

As a description from the perspective of the M2-brane, Basu and Harvey\wpcite{Basu:2004ed} suggested the following equation:
\begin{equation}\label{eq:BasuHarvey}
 \dder{x^6}X^\mu=\tfrac{1}{3!}\eps^{\mu\nu\kappa\lambda}[X^\nu,X^\kappa,X^\lambda]~,
\end{equation}
where the fields $X^\mu$ take values in some internal vector space endowed with some totally antisymmetric ternary bracket $[-,-,-]$; see appendix~\ref{app:A} for details. This equation has led to the development of the M2-brane models\wpcite{Bagger:2007jr,Gustavsson:2007vu,Aharony:2008ug}, and we also expect that it can teach us more about M5-branes. 

There are now two important problems to address. First, we should extend the self-dual string equation to the non-abelian setting. Second, we should try to establish a duality between the self-dual string equation and the Basu--Harvey equation analogous to the Nahm transform. The second problem would have very interesting mathematical implications. We will address the first problem in the following, and our discussion provides in principle a starting point for the second one. 

\subsection{Higher Lie algebras}\label{ssec:higher_lie}

An \wpem{N-manifold\/} $\CM$ is an $\NN$-graded manifold. That is, its algebra of functions or, more precisely, its structure sheaf is generated by elements of degree $0,1,2,\ldots$. These generators can be regarded as coordinates on $\CM$, and therefore $\CM$ consists of a \wpem{body\/} $M_0$ together with additional spaces fibered over the body:
\begin{equation}
 \CM\ = \ (M_0 \longleftarrow M_1 \longleftarrow M_2 \longleftarrow \ldots)~.
\end{equation}
Recall that by Batchelor's theorem\wpcite{JSTOR:1998201}, any supermanifold (i.e.\ $\RZ_2$-graded manifold) is diffeomorphic to a split supermanifold (i.e.\ a supermanifold whose $M_1$ is a vector bundle over $M_0$). This theorem can be extended to N-manifolds\wpcite{Bonavolonta:2012fh}, and we can therefore assume that $M_1\oplus M_2\oplus \ldots$ forms an $\NN$-graded vector bundle over $M_0$.

An \wpem{N$Q$-manifold\/} is an $\NN$-graded manifold together with a vector field $Q$ of degree 1, which satisfies $Q^2=0$. We call such a vector field a \wpem{homological vector field}. N$Q$-manifolds are known to physicists from BRST and BV-quantization as well as from closed string field theory, while mathematicians know them as differential graded algebras featuring, \wpeg, in the Chevalley--Eilenberg description of Lie algebras. 

Let us explain the latter in more detail. Consider an N-manifold concentrated in degree 1. That is, $\CM$ consists exclusively of a vector space $\frg$ and all vectors have degree 1. We also write $\CM=\frg[1]$. In terms of coordinates $\xi^\alpha$ of degree 1 on $\frg[1]$, a homological vector field is necessarily of the form
\begin{equation}\label{ex:Lie_algebra}
  Q=-\tfrac12 \xi^\alpha\xi^\beta f^\gamma_{\alpha\beta}\der{\xi^\gamma}~,
\end{equation}
where the $f^\gamma_{\alpha\beta}$ are some constants and the prefactor of $-\tfrac12$ is inserted for convenience. The identity $Q^2=0$ is equivalent to the Jacobi identity for a Lie bracket with structure constants $f^\gamma_{\alpha\beta}$.

\begin{exercise}
 Derive the Jacobi identity from $Q^2=0$.
\end{exercise}

We can now readily define \wpem{strong homotopy Lie algebras} or \wpem{$L_\infty$-algebras} and \wpem{$L_\infty$-algebroids}. An \wpem{$n$-term $L_\infty$-algebroid} is an N$Q$-manifold concentrated in degrees $0,\ldots,n$:
\begin{equation}
 \CM\ = \ (M_0 \longleftarrow M_1 \longleftarrow M_2 \longleftarrow \ldots \longleftarrow M_n \longleftarrow * \longleftarrow * \longleftarrow \ldots)~.
\end{equation}
Here, a $*$ denotes a one-point space, \wpie\ a 0-dimensional vector space. An \wpem{$n$-term $L_\infty$-algebra} is an N$Q$-manifold concentrated in degrees $1,\ldots,n$:
\begin{equation}
 \CM\ = \ (* \longleftarrow M_1 \longleftarrow M_2 \longleftarrow \ldots \longleftarrow M_n \longleftarrow * \longleftarrow * \longleftarrow \ldots)~.
\end{equation}
In particular, we saw in \eqref{ex:Lie_algebra} the example of a general $1$-term $L_\infty$-algebra and this is simply a Lie algebra. Baez \& Crans\wpcite{Baez:2003aa} have shown that 2-term $L_\infty$-algebras are categorically equivalent to semistrict Lie $2$-algebras. We can therefore be slightly sloppy and use the term $n$-term $L_\infty$-algebra and Lie $n$-algebra interchangeably. The former turn out to be very convenient for describing higher gauge algebras in higher gauge theories.

We are mostly interested in Lie 2-algebras, and therefore let us look at 2-term $L_\infty$-algebras in more detail. Here, we have the N$Q$-manifold
\begin{equation}
 \CM\ =\ (*\leftarrow \frg[1]\leftarrow \frh[2]\leftarrow * \leftarrow \ldots)~,
\end{equation}
and we use coordinates $\xi^\alpha$ and $\chi^\kappa$ of degree 1 and 2 on $\frg[1]$ and $\frh[2]$, respectively. The homological vector field is necessarily of the form
\begin{equation}\label{eq:hol_vec_field_2}
 Q=\pm m^\alpha_\kappa\chi^\kappa \der{\xi^\alpha}\pm \tfrac12 f^\gamma_{\alpha\beta}\xi^\alpha\xi^\beta\der{\xi^\gamma}\pm m_{\alpha\kappa}^\lambda\xi^\alpha\chi^\kappa\der{\chi^\lambda}\pm\tfrac{1}{3!}m^\kappa_{\alpha\beta\gamma}\xi^\alpha\xi^\beta\xi^\gamma\der{\chi^\kappa}~,
\end{equation}
and we shall address the correct signs later. On the shifted vector space $\sL=\CM[-1]=\frg[0]\oplus \frh[1]$, with graded basis $(\tau_\alpha,\sigma_\kappa)$, the structure constants contained in $Q$ induce higher products
\begin{equation}
 \mu_k:\sL^{\wedge k}\rightarrow \sL
\end{equation}
of degree $k-2$ according to
\begin{equation}\label{eq:higher_brackets}
\begin{aligned}
 \mu_1(\sigma_\kappa)&:=m^\alpha_\kappa\tau_\alpha~,\\
 \mu_2(\tau_\alpha,\tau_\beta):=f^\gamma_{\alpha\beta}\tau_\gamma~,~~~&\mu_2(\tau_\alpha,\sigma_\kappa):=m^\lambda_{\alpha\kappa}\sigma_\lambda~,\\
 \mu_3(\tau_\alpha,\tau_\beta,\tau_\gamma)&:=m^\kappa_{\alpha\beta\gamma}\sigma_\kappa~.
\end{aligned}
\end{equation}
Lie 2-algebras with trivial $\mu_3$ are called \wpem{strict Lie 2-algebras}.

The relation $Q^2$ amounts to the \wpem{homotopy Jacobi identity},\begin{equation}\label{eq:homotopyJacobi}
 \sum_{i+j=n}\sum_\sigma\chi(\sigma;\ell_1,\ldots,\ell_n)(-1)^{i\cdot j}\mu_{j+1}(\mu_i(\ell_{\sigma(1)},\cdots,\ell_{\sigma(i)}),\ell_{\sigma(i+1)},\cdots,\ell_{\sigma(i+j)})=0
\end{equation}
for all $\ell_i\in \sL$, where the sum runs over all \wpem{unshuffles} $\sigma$. These are permutations $\sigma$ with $\sigma(1)<\ldots<\sigma(i)$ and $\sigma(i+1)<\ldots< \sigma(i+j)$. The graded Koszul sign $\chi(\sigma;\ell_1,\cdots,\ell_n)$ is defined implicitly by the equation
\begin{equation}
 \ell_1\wedge \cdots \wedge \ell_n=\chi(\sigma;\ell_1,\cdots,\ell_n)\,\ell_{\sigma(1)}\wedge \cdots \wedge \ell_{\sigma(n)}~.
\end{equation}
That is, write the permutation $\sigma$ as a sequence of swaps of neighboring objects and count the number $s$ of such swaps involving at least one object of even degree. The Koszul sign is then $(-1)^s$.

\begin{exercise}
 Fix (some of) the signs in \eqref{eq:hol_vec_field_2} such that \eqref{eq:higher_brackets} is compatible with \eqref{eq:homotopyJacobi}.
\end{exercise}
Note that equivalently, we can invert the grading to a non-positive one, resulting in an $L_\infty$-algebra $\bar\sL=\frh[-1]\oplus \frg[0]$, in which the brackets $\mu_k$ carry degree $2-k$.

\subsection{Local higher gauge theory}\label{ssec:local_higher_gauge}

As a first step towards higher gauge theory, let us develop a local description of the necessary kinematical data. This involves the definition of higher gauge potential forms, curvature forms as well as the notion of infinitesimal gauge transformations over a contractible manifold $M$ for a given Lie 2-algebra $\sL$.

There is a natural equation on an $L_\infty$-algebra $\sL$, the so-called \wpem{homotopy Maurer--Cartan equation},
\begin{equation}\label{eq:hMC1}
  \sum_i \frac{(-1)^{k(k+1)/2}}{k!}\mu_k(\phi,\ldots,\phi)\ =\ 0~,
\end{equation}
An element $\phi\in \sL$ satisfying this equation is called a \wpem{Maurer--Cartan element}. Note that these equations are invariant under the infinitesimal gauge transformations
\begin{equation}\label{eq:hMCt1}
  \phi\rightarrow \phi+\delta \phi\ewith \delta \phi\ =\ \sum_k \frac{(-1)^{k(k-1)/2}}{(k-1)!}\mu_k(\gamma,\phi,\ldots,\phi)~,
\end{equation}
where $\gamma$ is a degree 0 element of $\sL$.

\begin{exercise}
 Verify the gauge invariance of \eqref{eq:hMC1} under \eqref{eq:hMCt1} in a simple case, e.g.\ when $\mu_k=0$ for $k\geq 3$.
\end{exercise}

In order to define curvatures, we have to combine the $L_\infty$-algebra $\sL$ with the differential graded algebra given by the de Rham complex on $M$, $(\Omega^\bullet(M),\dd)$. This is done by taking the tensor product of both algebras, which always carries a natural $L_\infty$-algebra structure\footnote{This holds actually for the tensor product of an arbitrary differential $\NN$-graded algebra and an $L_\infty$-algebra.}.

More precisely, we take the tensor product of $\Omega^\bullet(M)$ with that of $\bar \sL$, which has the inverted grading of $\sL$ and truncate to elements of positive degree, $\tilde\sL_{\geq 0}:=(\Omega^\bullet(M)\otimes \bar\sL)_{\geq 0}$. The total grading of an element of $\tilde \sL_{\geq 0}$ is the de Rham grading plus the grading in $\bar \sL$, $|\alpha\otimes \ell|=|\alpha|+|\ell|$. Explicitly, the vector subspace of degree $p$-elements for $p\geq 0$ is given by 
\begin{equation}
(\tilde\sL_{\geq 0})_p\ =\ (\Omega^p(M)\otimes \bar \sL_0)\oplus(\Omega^{p+1}(M)\otimes \bar \sL_{-1})~.
\end{equation}

For a tuple of elements $(\alpha_1\otimes\ell_1,\ldots,\alpha_k\otimes \ell_k)$ of $\tilde\sL_{\geq 0}$, the higher products $\tilde{\mu}_k$ read as
\begin{equation}
\tilde\mu_k(\alpha_1\otimes\ell_1,\ldots,\alpha_k\otimes \ell_k)\ =\ 
\left\{\begin{array}{ll}
(\dd \alpha_1) \otimes \ell_1+(-1)^{\deg(\alpha_1)}\alpha_1\otimes \mu_1(\ell_1)&\efor k=1~,\\
\pm\alpha_1\alpha_2\cdots \alpha_k\otimes \mu_k(\ell_1,\ldots,\ell_k)&\efor k>1~.\end{array}\right.
\end{equation}
Here, the $\mu_k$ are the higher products in $\bar L$, $\deg$ denotes the degrees in $\Omega^\bullet(M)$, and the sign $\pm$ in the case $k>1$ arises from moving graded elements of $\Omega^\bullet(M)$ past graded elements of $\sL$.

We can now consider Maurer--Cartan elements on $\tilde\sL_{\geq 0}$ and read off higher curvatures and infinitesimal gauge transformations. For an element $\phi=A-B$ of degree 1, where $A\in \Omega^1(M)\otimes \bar \sL_0$ and $B\in \Omega^2(M)\otimes \bar \sL_{-1}$, the homotopy Maurer--Cartan equation~\eqref{eq:hMC1} reads as 
\begin{equation}\label{eq:field_equations}
\begin{aligned}
\CF&\ :=\ \dd A+\tfrac{1}{2}\mu_2(A,A)-\mu_1(B)\ =\ 0~,\\
H&\ :=\ \dd B+\mu_2(A,B)-\tfrac{1}{3!}\mu_3(A,A,A)\ =\ 0~.
\end{aligned}
\end{equation}
Correspondingly, the infinitesimal gauge transformations parametrized by a degree 0 element $\gamma=\omega+\Lambda$ with $\omega\in\Omega^0(M)\otimes \bar \sL_0$ and $\Lambda\in \Omega^1(M)\otimes \bar \sL_{-1}$ are given by 
\begin{equation}\label{eq:gauge_trafos}
\begin{aligned}
  \delta A\ &=\ \dd \omega+\mu_2(A,\omega)-\mu_1(\Lambda)~,\\
  \delta B\ &=\ -\dd \Lambda-\mu_2(A,\Lambda)+\mu_2(B,\omega)+\tfrac{1}{2}\mu_3(\omega,A,A)~.
\end{aligned}
\end{equation}
In this way, we can construct the local higher curvatures and infinitesimal gauge transformations for higher gauge theory on any spacetime carrying a differential graded algebra and for any gauge $L_\infty$-algebra.

\begin{exercise}
 Derive formulas~\eqref{eq:field_equations} and \eqref{eq:gauge_trafos} from Equations~\eqref{eq:hMC1} and \eqref{eq:hMCt1}.
\end{exercise}

If one performs a detailed analysis of parallel transport via functors from higher path groupoids to the delooping of higher gauge groups\wpcite{Baez:2004in}, one finds that reparametri\-zation invariance of the surfaces and higher dimensional volumes involved requires all but the highest curvature form to vanish.

Being very optimistic, we can now postulate equations of motion for the gauge part of a theory of multiple M5-branes. The field content consists of a degree 1-element $A-B$ in $\tilde \sL_{\geq 0}$ for some gauge Lie 2-algebra $\sL$, which satisfies the equations
\begin{equation}
\begin{aligned}
 H&:=\dd B+\mu_2(A,B)-\tfrac{1}{3!}\mu_3(A,A,A)=*H~,\\
 \CF&:=\dd A+\mu_2(A,A)-\mu_1(B)=0~.
\end{aligned}
\end{equation}
Note that the additional degrees of freedom contained in the one-form potential are fully determined by the equation $\CF=0$.

\subsection{Further reading}

The holonomy functor is explained in great detail in Baez \& Huerta\wpcite{Baez:2010ya} and Baez \& Schrei\-ber\wpcite{Baez:2004in}. A detailed discussion of self-dual strings and the duality can be found \wpeg\ in section 3 of my paper\wpcite{Saemann:2010cp}. N$Q$-manifolds are thoroughly introduced in Roytenberg\wpcite{Roytenberg:0203110}. Their relation to $L_\infty$-algebras is reviewed in the papers\wpcite{Lada:1992wc,Lada:1994mn}, where also the homotopy Maurer--Cartan equations and their infinitesimal gauge symmetries are found. For a discussion of this in the context of string field theory see Zwiebach\wpcite{Zwiebach:1992ie}. 

The construction of local higher gauge theory as done in the previous section was first given in\wpcite{Jurco:2014mva}. I chose to follow this route, because it is the shortest way of deriving local higher gauge theory that I am aware of. A more geometrical approach involving morphisms of N$Q$-manifolds arises from a local version\wpcite{Bojowald:0406445,Sati:0801.3480,Kotov:2010wr,Gruetzmann:2014ica} of a construction of Atiyah\wpcite{Atiyah:1957}; see also Section 2 of the paper\wpcite{Ritter:2015zur} for a concise review and a further extension of this description of local higher gauge theory.

A particularly impressive demonstration of the usefulness of higher Lie algebras in physics is the reproduction of the complete brane scan in type II superstring theory and M-theory from considering cocycle extensions of super $L_\infty$-algebras\wpcite{Fiorenza:2013nha}.

\section{Categorification}\label{sec:3}

Let us now come to the mathematical concepts which will allow us to turn our notion of local higher gauge theory into a global one. For simplicity, we shall focus on strict 2-categories, strict 2-groups and strict Lie 2-algebras. A more general picture based on weak 2-categories (which are also known as bicategories) has also been worked out\wpcite{Jurco:2014mva}. 

\subsection{(Strict) 2-categories}

Formally, a 2-category is a category enriched over $\CatCat$. More explicitly, the idea here is to have objects (points), morphisms (oriented lines) and morphisms between morphisms (oriented surfaces):
\begin{equation}\label{diag:2-category}
 \myxymatrix{ a
&& b
  \ar@/_4ex/[ll]_{f}="g1"
  \ar@{}[ll]|{}="g2"
  \ar@/^4ex/[ll]^{g}="g3"
  \ar@{=>}^{\alpha} "g1"+<0ex,-2ex>;"g3"+<0ex,2ex>
}
\end{equation}
A strict 2-category $\CCC$ consists of a set\footnote{For simplicity, we restrict ourselves to small categories based on sets instead of classes.} of \wpem{objects} $\CCC_0$, denoted $a,b,c,\ldots$ and for each pair of objects $(a,b)$ a category $\CCC(a,b)$ of morphisms. This category, in turn, contains objects, called \wpem{1-morphisms} $f:a\rightarrow b$, and morphisms, called \wpem{2-morphisms} $\alpha:f\Rightarrow g$. The composition $\circ$ in $\CCC(a,b)$ is known as \wpem{vertical composition}, as the composed 2-morphisms are vertically composed in diagrams such as \eqref{diag:interchange_law}. There is also a functor $\otimes:\CCC(a,b)\times \CCC(b,c)\rightarrow \CCC(a,c)$, known as \wpem{horizontal composition}. Everything is unital and associative, and we automatically get the interchange law
\begin{equation}
 (\beta'\circ \beta)\otimes (\alpha'\circ \alpha)=(\beta'\otimes \alpha')\circ(\beta\otimes \alpha)~,
\end{equation}
\wpcf\ \eqref{eq:interchange_law}.

Just as the category $\CatSet$ consisting of sets and morphisms between sets is the ``mother of all categories,'' the 2-category $\CatCat$ consisting of categories, functors and natural transformations is the mother of all 2-categories.

To define 2-functors, we note that the ordinary definition is not quite sufficient for our purposes, and we need to generalize to \wpem{pseudofunctors}. Such a pseudofunctor between two 2-categories $\CCC$ and $\CCD$ is given by
\begin{itemize}
 \item a function $\Phi_0:\CCC_0\rightarrow \CCD_0$,
 \item a functor $\Phi_{1}^{ab}:\CCC(a,b)\rightarrow \CCD(\Phi_0(a),\Phi_0(b))$,
 \item a 2-morphisms $\Phi_{2}^{abc}:\Phi_{1}^{ab}(f)\otimes_\CCD \Phi_{1}^{bc}(g)\Rightarrow \Phi_{1}^{ac}(f\otimes_\CCC g)$,
 \item a 2-morphism $\Phi_{2}^{a}:\id_{\Phi_0(a)}\Rightarrow \Phi_{1}^{aa}(\id_a)$.
\end{itemize}
The last two 2-morphisms are responsible for the prefix `pseudo.' It will turn out that we can restrict ourselves to \wpem{normalized pseudofunctors}, \wpie\ pseudofunctors with $\Phi_{2}^{a}$ the identity, without loss of generality. We still have a compatibility relation for the 2-cells given by $\Phi_{2}^{abc}$, which arises from the diagram
\begin{equation}
\xymatrixcolsep{3pc}
\myxymatrix{
& \cdots \ar@{=>}[dr] & \\
 (\Phi_1^{ab}(x)\,\tilde \otimes\,\Phi_1^{bc}(y))\,\tilde \otimes\, \Phi_1^{cd}(z) \ar@{=>}[ur]^{\Phi_2^{abc}\otimes \id}  \ar@{=>}[d]_{=} & & \Phi_1^{ad}((x\otimes y)\otimes z) \ar@{=>}[d]_{=}\\
 \Phi_1^{ab}(x)\,\tilde \otimes\,(\Phi_1^{bc}(y)\,\tilde \otimes\, \Phi_1^{cd}(z))\ar@{=>}[dr] &  & \Phi_1^{ad}(x\otimes(y\otimes z))\\
 & \cdots \ar@{=>}[ur] & }
\end{equation}

\begin{exercise}
 Label the arrows and fill in the $\cdots$. From the commutativity of the diagram, write down the equation satisfied by the $\Phi_2^{abc}$. The answer for weak 2-categories, which reduce for trivial associators and unitors to the strict case, is found in the literature\wpcite{Jurco:2014mva}.
\end{exercise}

Analogously, one defines natural 2-transformations\wpcite{Jurco:2014mva}.

\subsection{Strict 2-groups}

The first ingredient in the definition of a principal bundle is a structure group, and we therefore need to find a higher analogue. Note that any group $\sG$ gives rise to a category $\sB\sG\rightrightarrows *$, where source and target are trivial, $\id_*=\unit_\sG$ and composition is given by group multiplication. This category is special, as all morphisms have an inverse. Such categories are called \wpem{groupoids}.

\begin{exercise}
 Briefly convince yourself that the morphisms of any groupoid with a single object form a group.
\end{exercise}

Correspondingly, we would like to define a 2-group $\CCG$ as a 2-groupoid $\sB\CCG$ with a single object. That is, we have a 2-category with a single object and invertible 1- and 2-morphisms:
\begin{equation}
 \sB\CCG:= (* \leftleftarrows \CCG_0 \leftleftarrows \CCG_1)~.
\end{equation}
Inversely, $\CCG$ is the morphism category in $\sB\CCG$ over $*$,
\begin{equation}
 \CCG=(\CCG_0\leftleftarrows \CCG_1)=\sB\CCG(*,*)~.
\end{equation}
This yields indeed the definition of a strict 2-group.

It has been shown\wpcite{Baez:0307200} that these strict 2-groups are categorically equivalent to \wpem{crossed modules of groups}. The latter consist of a pairs of groups $\sH,\sG$ together with homomorphisms $\dpar: \sH\rightarrow \sG$ and actions $\acton:G\ltimes \sH\rightarrow \sH$ satisfying
\begin{equation}
 \dpar(g\acton h)=g\acton \dpar(h)~,~~\dpar(h_1)\acton h_2=h_1h_2h_1^{-1}
\end{equation}
for all $g\in \sG$ and $h,h_{1,2}\in \sH$. To reconstruct the corresponding strict 2-group, put
\begin{equation}
 \begin{aligned}
  &\CCG_0=\sG~,~~~\CCG_1=\sG\ltimes \sH~,~~~\sfs(g,h)=g~,~~~\sft(g,h)=\dpar(h)g~,~~~\id(g)=(g,\unit)~,\\
  &\hspace{1cm}g_1\otimes g_2=g_1g_2~,~~~(g_1,h_1)\otimes(g_2,h_2)=(g_1g_2,h_1(g_1\acton h_2))~,\\
  &\hspace{3.5cm}(\dpar(h_1)g,h_2)\circ (g,h_1)=(g,h_2h_1)~.
 \end{aligned}
\end{equation}
Conversely, a crossed module of groups is derived from the strict 2-group $\CCG$ by putting\footnote{look up ``Moore complex''} $\sH=\ker(\sfs)$ and $\sG=\CCG_0$.

\begin{exercise}
Complete the inverse map. The solution is found in Baez \& Lauda\wpcite{Baez:0307200}.
\end{exercise}

\subsection{Principal bundles as functors}

We now come to the description of principal bundles from the $n$-POV, which goes back to Segal\wpcite{Segal1968} and which is suitable for categorification. Recall that the \v Cech groupoid $\check \CCC(Y)$ of a surjective submersion $\pi:Y\twoheadrightarrow M$ has objects $Y$ and morphisms 
\begin{equation}
 Y^{[2]}=Y\times_M Y:=\{~(y_1,y_2)~|~\pi(y_1)=\pi(y_2)~\}
\end{equation}
with obvious structure maps. In the case of an ordinary cover $Y=\sqcup_i U_i$, the objects are pairs $(x,i)$ with $x\in U_i$ and the morphisms are triples $(x,i,j)$, $x\in U_i\cap U_j$. We have $\sfs(x,i,j)=(x,j)$ and $\sft(x,i,j)=(x,i)$ as well as $\id(x,i)=(x,i,i)$, $(x,i,j)\circ (x,j,k)=(x,i,k)$ and $(x,i,j)^{\circ-1} =(x,j,i)$. This groupoid encodes all necessary information about the manifold $M$.\footnote{Each manifold $M$ gives rise to a trivial groupoid $M\rightrightarrows M$. This groupoid is Morita equivalent (\wpie\ equivalent in the 2-category of presentable stacks) to any \v Cech groupoid $\check \CCC(Y)$ arising from a surjective submersion $Y\twoheadrightarrow M$.} Note that the \v Cech groupoid can trivially be regarded as a 2-groupoid by adding all identity 2-morphisms. That is, the corresponding 2-category has objects $Y$ and the categories of morphisms combine to the trivial category $Y^{[2]}\rightrightarrows Y^{[2]}$.

On the other hand, we need to choose a Lie group $\sG$, which forms the structure or gauge group of our principal bundle. To put it on equal footing with the \v Cech groupoid, we immediately switch to the groupoid $\sB\sG:=\sG\rightrightarrows *$. 

We now define a principal bundle subordinate to the surjective submersion $Y\twoheadrightarrow M$ as a functor $\check \CCC(Y)\rightarrow \sB\sG$. We have the following diagram
\begin{equation}
 \xymatrixcolsep{5pc}
\myxymatrix{
 \{(x,i,j)\}\ar@{->}[r]^{\{g_{ij}(x)\}} \ar@<-.5ex>[d] \ar@<.5ex>[d] & \sG \ar@<-.5ex>[d] \ar@<.5ex>[d]\\
 \{(x,i)\} \ar@{->}[r] & {*}
}
\end{equation}
where the compatibility with the identity and composition contained in the definition of a functor implies $g_{ii}=\id_*=\unit_{\sG}$ on $U_i=U_{ii}$ and $g_{ij}(x)g_{jk}(x)=g_{ik}(x)$ on $U_{ijk}$, respectively.

A bundle isomorphism is accordingly given by natural transformations, which are encoded in maps $\gamma_i:U_i\rightarrow \sG$ such that the following diagram commutes:
\begin{equation}
 \xymatrixcolsep{5pc}
\myxymatrix{
 {*} \ar@{->}[d]_{\gamma_i(x)} & {*} \ar@{->}[d]^{\gamma_j(x)}\ar@{->}[l]_{\tilde g_{ij}(x)}\\
 {*}  & {*}\ar@{->}[l]_{g_{ij}(x)}
}
\end{equation}
We arrive at the cocycle relation $\gamma_i(x)\tilde g_{ij}(x)=g_{ij}(x)\gamma_j(x)$ and altogether, we have recovered the first \v Cech cohomology class with values in the sheaf of smooth $\sG$-valued functions, \wpcf~Equations \eqref{eq:1-cocycle} and \eqref{eq:1-coboundary}.

\subsection{Principal 2-bundles}\label{ssec:principal_2_bundles}

We now have everything at our disposal to define principal 2-bundles: A \wpem{principal 2-bundle} over a manifold $M$ subordinate to a cover $Y\twoheadrightarrow M$ with structure 2-group $\CCG$ is a normalized pseudofunctor from the \v Cech 2-groupoid $\check \CCC(Y)$ to the 2-groupoid $\sB\CCG$. If $\sH\stackrel{\dpar}{\rightarrow} \sG$ is the crossed module of Lie groups corresponding to $\CCG$, the cocycle resulting from this definition is encoded in functions $g_{ij}:U_{ij}\rightarrow \sG$ and $h_{ijk}:U_{ijk}\rightarrow \sH$ satisfying 
\begin{equation}\label{eq:C_cocycles}
 \begin{aligned}
  \dpar(h_{ijk})g_{ij}g_{jk}&=g_{ik}~,\\
  h_{ikl}h_{ijk}&=h_{ijl}(g_{ij}\acton h_{jkl})~.
 \end{aligned}
\end{equation}

Isomorphisms of principal 2-bundles are natural 2-transformations between the corresponding pseudofunctors.

\begin{exercise}
Verify the cocycle condition~\eqref{eq:C_cocycles} and derive the corresponding coboundary relations. Then compare your results to the answer for weak 2-groups\wpcite{Jurco:2014mva}.
\end{exercise}

Among important examples of principal 2-bundles, we have ordinary, principal $\sG$-bundles in the case of a crossed module $*\stackrel{\dpar}{\rightarrow} \sG$, abelian gerbes in the case of a crossed module $\sU(1)\stackrel{\dpar}{\rightarrow}*$ and twisted principal $\sG$-bundles in the case of a crossed module $\sU(1)\stackrel{\dpar}{\rightarrow} \sG$. Thus we note that principal 2-bundles nicely unify non-abelian principal bundles and abelian gerbes.

To add categorified connections to our principal 2-bundles, we have to discuss Lie 2-algebras and how they are obtained by differentiating Lie 2-groups.

\subsection{Differentiating Lie 2-groups}\label{ssec:differenitation}

An integration of $L_\infty$-algebras can be performed\wpcite{Henriques:2006aa}, but the procedure is very cumbersome. As always, differentiation is easier than integration, and we therefore start with a Lie 2-group $\CCG$. An $n$-POV on the Lie algebra $\sLie(\sG)$ of a Lie group $\sG$ was suggested by \v Severa\wpcite{Severa:2006aa}. In this picture, we consider the functor that maps supermanifolds $X$ to descent data for principal $\sG$-bundles subordinate to the surjective submersions $X\times \FR^{0|1}\rightarrow X$. As a vector space, the Lie algebra is recovered as the moduli space of such functors. Moreover, its Chevalley--Eilenberg differential is obtained as the action of one of the generators of $\sHom(\FR^{0|1},\FR^{0|1})$ on this moduli space. This description readily categorifies and in his paper\wpcite{Severa:2006aa}, \v Severa discusses the differentiation of what one calls $(\infty,1)$-groups.

Let us discuss this construction for Lie groups in detail. Descent data for a principal $\sG$-bundle subordinate to the surjective submersion $Y=X\times \FR^{0|1}\rightarrow X$ is captured by functions $g$ from the morphisms of the \v Cech groupoid $\check\CCC(Y)$ given by $X\times \FR^{0|2}$ to $\sG$ such that
\begin{equation}\label{eq:cocycle}
 g(\theta_0,\theta_1,x)g(\theta_1,\theta_2,x)=g(\theta_0,\theta_2,x)~,
\end{equation}
where $\theta_{0,1,2}\in \FR^{0|1}$ and $x\in X$. This equation immediately implies that
\begin{equation}
 g(\theta,\theta,x)=1\eand g(\theta_1,\theta_2,x)=(g(\theta_2,\theta_1,x))^{-1}~.
\end{equation}
Putting $\theta_1$ to $0$ and renaming $\theta_2$ to $\theta_1$ in \eqref{eq:cocycle}, we therefore have
\begin{equation}
 g(\theta_0,\theta_1,x)=g(\theta_0,0,x)(g(\theta_1,0,x))^{-1}~.
\end{equation}
Fixing the parametrization\footnote{By this sum, we mean the obvious one involving the local diffeomorphism between $\sG$ and $T_\unit\sG$.}
\begin{equation}
 g(\theta_0,0,x)=\unit+a\theta_0~,
\end{equation}
where $a\in T_\unit\sG[1]=\frg[1]$, we can compute
\begin{equation}
 g(\theta_0,\theta_1)=\unit+a(\theta_0-\theta_1)+\tfrac12 [a,a]\theta_0\theta_1~.
\end{equation}
Moreover, we have the following natural vector field $Q$ acting on $g(\theta_0,\theta_1,x)$:
\begin{equation}
 Q g(\theta_0,\theta_1,x):=\dder{\eps}g(\theta_0+\eps,\theta_1+\eps,x)~,
\end{equation}
which induces the action
\begin{equation}
 Q a=-\tfrac12[a,a]~~~\mbox{or}~~~Q a^\alpha=-\tfrac12 f^\alpha_{\beta\gamma}a^\beta a^\gamma
\end{equation}
for $a=a^\alpha\tau_\alpha$ in some basis $\tau_\alpha$ of $\frg$. Altogether, we recovered the Lie algebra in the form of an N$Q$-manifold, as described in Section~\ref{ssec:higher_lie}.

If we now apply this procedure to a crossed module of Lie groups written as a strict Lie 2-group, we obtain a crossed module of Lie algebras.

\begin{exercise}
 Construct analogously the Lie 2-algebra of a strict Lie 2-group $\CCG=(\sG\ltimes \sH\rightrightarrows\sG)$. If you should get stuck, you can compare to the more general computation in the weak case\wpcite{Jurco:2014mva}.
\end{exercise}

In our paper\wpcite{Jurco:2014mva}, we pushed the analysis of \v Severa further and considered equivalences between the functors to descent data. This induces isomorphisms on the moduli, and in the case of an ordinary Lie group, we obtain
\begin{equation}
 a\mapsto \tilde a=\gamma^{-1}a\gamma+\gamma^{-1}Q\gamma~,
\end{equation}
where $\gamma\in \sG$. Replacing $Q$ with the de Rham differential, we recover the finite gauge transformations, \wpcf~Equation \eqref{eq:1-coboundary}.

\begin{exercise}
 Derive analogously the finite gauge transformations for local higher gauge potentials for a strict Lie 2-group $\CCG=(\sG\ltimes \sH\rightrightarrows\sG)$. Again, the computation in a more general case has been spelled out\wpcite{Jurco:2014mva} and in this paper, the results for the strict case are listed separately.
\end{exercise}

\subsection{Summary of the construction}

The construction given above readily generalizes to the extent that a higher gauge structure can be defined. Given an arbitrarily general spacetime\footnote{Note that $M$ does not have to be a manifold, it can also be a categorical space, a Lie groupoid (\wpeg\ describing an orbifold) or a higher Lie $n$-groupoid\wpcite{Ritter:2015zur,Jurco:2016qwv}.} $M$ and a general gauge groupoid, our constructions produce the kinematical data for the corresponding higher gauge field theories. These can be higher gauge theories or higher gauged sigma models. We first construct the higher principal bundle as in Section~\ref{ssec:principal_2_bundles}. Next, we derive the gauge algebra as in Section~\ref{ssec:differenitation}. We then define the local connective structure along the lines of Section~\ref{ssec:local_higher_gauge} and glue all fields together with the finite gauge transformations derived as in Section~\ref{ssec:differenitation}.

In the case of a strict Lie 2-group $\CCG=(\sG\ltimes \sH\rightrightarrows\sG)$, this yields the following non-abelian Deligne cocycle subordinate to a cover $\sqcup U_i$. A cochain consists of forms
\begin{equation}
 \begin{aligned}
  g_{ij}&\in \Omega^0(U_{ij},\sG)~,~~~&A_i&\in\Omega^1(U_i,\sLie(\sG))~,~~~&B_i&\in\Omega^2(U_i,\sLie(\sH))~,\\
  h_{ijk}&\in \Omega^0(U_{ijk},\sH)~,~~~&\Lambda_{ij}&\in \Omega^1(U_{ij},\sLie(\sH))
 \end{aligned}
\end{equation}
satisfying the cocycle relations
\begin{equation}
 \begin{aligned}
  \dpar(h_{ijk})g_{ij}g_{jk}&\ =\ g_{ik}\eand h_{ikl}h_{ijk}\ =\ h_{ijl}(g_{ij}\acton h_{jkl})~,\\
 A_j&\ =\ g^{-1}_{ij} A_i g_{ij}+g^{-1}_{ij} \dd g_{ij}-\dpar(\Lambda_{ij})~,\\
 B_j&\ = \ g^{-1}_{ij}\acton B_i -A_j\acton\Lambda_{ij}-\dd \Lambda_{ij}-\Lambda_{ij}\wedge \Lambda_{ij}~,\\
  \Lambda_{ik}&\ =\ \Lambda_{jk}+g_{jk}^{-1}\acton\Lambda_{ij}-g_{ik}^{-1}\acton(h_{ijk}\nabla_ih_{ijk}^{-1})~.
 \end{aligned}
\end{equation}
The corresponding curvatures read as
\begin{equation}
  \CF_i\ :=\ \dd A_i+\tfrac12[A_i,A_i]-\dpar(B_i)\eand H_i\ :=\ \nabla B_i\ :=\ \dd B_i+A_i\acton B_i~.
\end{equation}

\begin{exercise}
 Write down the corresponding coboundary relations between two cocycles $(g,h,A,B,\Lambda)$ and $(\tilde g,\tilde h,\tilde A,\tilde B,\tilde \Lambda)$.
\end{exercise}

\subsection{Further reading}

The first non-abelian higher gerbes were defined by Breen \& Messing\wpcite{Breen:math0106083,Baez:2002jn} which were then generalized in various papers\wpcite{Aschieri:2003mw,Aschieri:2004yz,Bartels:2004aa,Baez:2004in,Jurco:2005qj,Schreiber:2008aa,Jurco:2016qwv}. For the general understanding, it is also very helpful to read up on gerbes\wpcite{Giraud:1971,0817647309}, particularly in the form of Murray's bundle gerbes\wpcite{Murray:9407015,Murray:2007ps}.

Higher gauge theory was probably first  studied by Baez\wpcite{Baez:2002jn} and Baez \& Schreiber\wpcite{Baez:2004in,Baez:0511710}. A very general and useful framework for describing higher groupoids are simplicial sets forming Kan complexes, and the corresponding notion of higher gauge theory can be found in our paper\wpcite{Jurco:2016qwv}. Particularly important examples of Lie 2-groups are the 2-group models of the String group, a higher version of the spin group. Higher gauge theory with these 2-groups has also been developed\wpcite{Demessie:2016ieh} and the underlying description involves the weak 2-category of bibundles which is the 2-category of presentable stacks mentioned above.

A very general framework for studying differential cohomology has been developed by Schreiber\wpcite{Schreiber:2013pra}, which subsumes our above constructions. 

\section{Constructing (2,0)-theories}\label{sec:4}

Let us now come to an application of our above framework, demonstrating its usefulness. In the following, we summarize the construction of $\CN=(2,0)$-theories using principal 2-bundles over twistor spaces\wpcite{Saemann:2012uq}.

\subsection{Twistors}

Twistors were proposed in 1967 by Penrose as a path to quantum gravity. From quantum mechanics, they inherit complex geometry and non-locality, while from general relativity, they inherit a relation to light rays and null spaces. Originally, twistor space was defined as the space of light cones. Given a point $x\in \FR^{1,3}$, the backwards light cone, intersected by the hypersurface $x^0=-1$ looks like a sphere: $(x^1)^2+(x^2)^2+(x^3)^2=1$. We can therefore identify twistor space with $\FR^{1,3}\times S^2$.

Twistor spaces find applications in classical integrable field theories, describing their solution spaces. Moreover, various approaches to computing scattering amplitudes are based on twistor spaces. Here, we focus on the former. For a comprehensive summary, see Wolf's review\wpcite{Wolf:2010av}.

Consider the instanton equation on $\FR^4$, $F=*F$, where $F$ is the curvature of the non-abelian connection on a principal $\sG$-bundle $P$.\footnote{An actual instanton is encoded in a gauge potential satisfying certain fall-off conditions so that the underlying principal bundle $P$ effectively becomes a bundle over the compactification $S^4$ of $\FR^4$. Otherwise, the bundle $P$ would necessarily be trivial and so would the instantons, which describe the topology of $P$. Physically, this is done by demanding that the Yang--Mills action functional is finite when evaluated on instanton solutions.} It turns out that it is convenient to work in the complex case $\FC^4$. In principle, reality conditions can be imposed at each step in our construction to recover the real case. Also, it is very helpful to switch to spinor notation,
\begin{equation}
 x^{\alpha\ald}=x^\mu\sigma_\mu^{\alpha\ald}=
 \left(\begin{array}{cc} 
        x^1+\di x^2 & x^3+\di x^4 \\
        -x^3+\di x^4 & x^1-\di x^2
       \end{array}\right)~,~~~\dpar_{\alpha\ald}x^{\beta\bed}=\delta_\alpha^\beta\delta_\ald^\bed~,~~~|x|^2=\det(x^{\alpha\ald})~.
\end{equation}
The curvature $F$ then splits up into components
\begin{equation}
 F_{\alpha\ald,\beta\bed}=\dpar_{\alpha\ald}A_{\beta\bed}-\dpar_{\beta\bed}A_{\alpha\ald}+[A_{\alpha\ald},A_{\beta\bed}]=\eps_{\alpha\beta}f_{\ald\bed}+\eps_{\ald\bed}f_{\alpha\beta}~,
\end{equation}
where $f_{\alpha\beta}$ contains the self-dual part of $F$, while $f_{\ald\bed}$ contains the anti-self-dual part of $F$. The self-duality equation therefore reduces to $f_{\ald\bed}=0$ or
\begin{equation}
 \lambda^\ald\lambda^\bed F_{\alpha\ald,\beta\bed}=0
\end{equation}
for all commuting spinors $\lambda^\ald$. This equation scales homogeneously in the commuting spinor, and we can therefore regard $\lambda_\ald=\eps_{\ald\bed}\lambda^\bed$ as homogeneous coordinates on $\CPP^1$.\footnote{We can avoid discussing patches by working in homogeneous coordinates over $\CPP^1\cong S^2$. To do so consistently, we simply have to ensure that all functions and sections have the appropriate homogeneous power in these coordinates.} The latter parametrize so-called \wpem{$\alpha$-planes} in $\FC^4$, that is, self-dual null-planes:
\begin{equation}
 x^{\alpha\ald}=x^{\alpha\ald}_0+\kappa^\alpha\lambda^\ald~,
\end{equation}
where $\kappa^\alpha$ is arbitrary. These planes are null in the sense that $|x^{\alpha\ald}-x_0^{\alpha\ald}|=0$. If we now factor out the dependence of $\alpha$-planes on the base point $x_0^{\alpha\ald}$, we obtain the following double fibration:
\begin{equation}\label{eq:DoubleFibration}
 \begin{picture}(50,40)
  \put(0.0,0.0){\makebox(0,0)[c]{$\CT^3$}}
  \put(64.0,0.0){\makebox(0,0)[c]{$\FC^4$}}
  \put(34.0,33.0){\makebox(0,0)[c]{$\FC^4\times \CPP^1$}}
  \put(7.0,18.0){\makebox(0,0)[c]{$\pi_1$}}
  \put(55.0,18.0){\makebox(0,0)[c]{$\pi_2$}}
  \put(25.0,25.0){\vector(-1,-1){18}}
  \put(37.0,25.0){\vector(1,-1){18}}
 \end{picture}
\end{equation}
We have coordinates $(x^{\alpha\ald},\lambda_\ald)$ on $\FC^4\times \CPP^1$ and coordinates $(z^\alpha,\lambda_\ald)$ on $\CT^3$, where the projection $\pi_2$ is trivial and $\pi_1$ is given by
\begin{equation}
 \pi_1(x^{\alpha\ald},\lambda_\ald)=(z^\alpha,\lambda_\ald):=(x^{\alpha\ald}\lambda_\ald,\lambda_\ald)~.
\end{equation}
We see that $\CT^3$ is a rank 2 vector bundle over $\CPP^1$ and its sections are homogeneous polynomials of degree 1. That is, $\CT^3$ is the total space of the vector bundle $\CO(1)\oplus \CO(1)\rightarrow \CPP^1$, which is diffeomorphic as a real manifold to the space $\FR^{1,3}\times S^2$ we introduced above as twistor space. The manifold $\CT^3$ can be covered by two patches $\hat U_+$ and $\hat U_-$, which are preimages of two patches $U_+$ and $U_-$ covering the sphere under the vector bundle projection.

Finally, note that the holomorphic vector fields in $T(\FC^4\times \CPP^1)$ along the fibration $\pi_1$ are linear combinations of
\begin{equation}\label{eq:defV}
 V_\alpha=\lambda^\ald\dpar_{\alpha\ald}~,
\end{equation}
since $V_\alpha z^\beta=\delta_\alpha^\beta \lambda^\ald\lambda_\ald=\delta_\alpha^\beta \eps^{\ald\bed}\lambda_\bed\lambda_\ald=0$ and $V_\alpha\lambda_\ald=0$.

\subsection{Solutions to integrable field equations}

Let us put a topologically trivial holomorphic principal $\sG$-bundle $\hat P$ over $\CT^3$, which becomes holomorphically trivial on every $\CPP^1$ embedded into $\CT^3$. The latter condition is rather technical and implies that the associated vector bundle for the fundamental representation of the gauge group has trivial first Chern class. Such a bundle $\hat P$ is described by a transition function $g_{+-}$ on $\hat U_+\cap \hat U_-$. Note that the preimages $U'_{\pm}$ of the patches $\hat U_\pm$ along $\pi_1$ cover $\FC^4\times \CPP^1$. Therefore, the pullback of $\hat P$ along $\pi_1$ has transition function $\pi_1^*g_{+-}$ on $U'_+\cap U'_-$, which satisfies
\begin{equation}
 V_\alpha \pi_1^*g_{+-}=0\eand \pi_1^*g_{+-}=\gamma_+^{-1}\gamma_-~,
\end{equation}
where $\gamma_{\pm}$ are holomorphic $\sG$-valued functions on $U'_{\pm}$. The first equation is a consequence of the pullback, the second results from $\hat P$ being holomorphically trivial on each $\CPP^1\embd \CT^3$. We then have a global 1-form\footnote{This is actually an element of the complex of relative differential forms, as explained in detail \wpeg\ in Ward \& Wells\wpcite{Ward:1990vs}.}
\begin{equation}
 A_\alpha:=\psi_+V_\alpha \psi_+^{-1}=\psi_-V_\alpha\psi_-^{-1}
\end{equation}
with
\begin{equation}\label{eq:lin_sys}
 (V_\alpha+A_\alpha)\psi_\pm=0~.
\end{equation}
Since the vector fields $V_\alpha$, which were defined in \eqref{eq:defV}, are linear in $\lambda_\ald$ and form global objects dual to global 1-forms $e^\alpha$, $A_\alpha$ are the components of a global 1-form $A=A_\alpha e^\alpha$, which is also linear in $\lambda_\ald$: $A_\alpha=\lambda^\ald A_{\alpha\ald}$. The compatibility condition of the linear system~\eqref{eq:lin_sys}, which is the necessary condition for a solution to exist, reads as
\begin{equation}
 [V_\alpha+A_\alpha,V_\beta+A_\beta]=0~~~\mbox{or}~~~\lambda^\ald\lambda^\bed[\dpar_{\alpha\ald}+A_{\alpha\ald},\dpar_{\beta\bed}+A_{\beta\bed}]=0~,
\end{equation}
where $A_{\alpha\ald}$ are the components of a gauge potential on $\FC^4$. We know that $\psi_\pm$ is a solution to \eqref{eq:lin_sys}. Therefore, the gauge potential $A_{\alpha\ald}$ defines a connection 1-form for an instanton on $\FC^4$. The resulting map, which takes a holomorphic principal bundle over $\CT^3$ to a self-dual connection on $\FR^4$ is known as the \wpem{Penrose--Ward transform}, and it is one direction of the following general theorem\wpcite{Ward:1977ta}:
\begin{theorem}\label{thm:Ward}
 Topologically trivial principal bundles over $\CT^3$ which become holomorphically trivial when restricted to any $\CPP^1\embd \CT^3$ are in one-to-one correspondence with instanton solutions on $\FC^4$, modulo isomorphisms on both sides.
\end{theorem}
One can prove this theorem by performing the (obvious) inverse construction and showing that post- and pre-composition with the original construction yields two identity maps. Note that the inverse construction involves a non-abelian Poincar\'e lemma for relative connections.

\subsection{Twistor space for self-dual 3-forms}

It turns out that a similar description to the one of instantons given above also exists for $\CN=4$ super Yang--Mills theory in four dimensions. It is therefore an obvious question whether we can find a twistor space for self-dual 3-forms, which we might then want to supersymmetrically extend to derive a non-abelian (2,0)-theory. This is indeed possible, and we sketch the construction in the following.

Let us describe $\FC^6$ again in spinor coordinates
\begin{equation}
 x^{AB}=-x^{BA}:=\sigma_M^{AB}x^M=\left(\begin{array}{cccc}
                        0 & x^0+x^5 & -x^3-\di x^4 & -x^1+\di x^2\\
			-x^0- x^5 & 0 & -x^1-\di x^2 & x^3-\di x^4\\
			x^3+\di x^4 & x^1+\di x^2 & 0 & -x^0+x^5\\
			x^1-\di x^2 & -x^3+\di x^4 & x^0-x^5 & 0
                       \end{array}\right),
\end{equation}
where $A=1,\ldots,4$, with
\begin{equation}                       
  x_{AB}:=\tfrac12\eps_{ABCD}x^{CD}~,~~~|x|^2=\det(x^{AB})~.
\end{equation}
A 1-form in spinor notation has components $A_{AB}=-A_{BA}$, a 2-form has components $B^A{}_B$ with vanishing trace: $B^A{}_A=0$ and a 3-form splits into two components $(H^{AB}=H^{BA},H_{AB}=H_{BA})$, where the first one is the anti-self-dual part and the second one is the self-dual part. The self-duality equation therefore reads as
\begin{equation}
 H^{AB}\lambda_A\lambda_B=0~,
\end{equation}
where $\lambda_A$ is a homogeneous coordinate on $\CPP^3$, parameterizing self-dual $\alpha$-planes in $\FC^6$. Correspondingly, we have the double fibration
\begin{equation}\label{eq:DoubleFibration2}
 \begin{picture}(50,40)
  \put(0.0,0.0){\makebox(0,0)[c]{$\CT^6$}}
  \put(64.0,0.0){\makebox(0,0)[c]{$\FC^6$}}
  \put(34.0,33.0){\makebox(0,0)[c]{$\FC^6\times \CPP^3$}}
  \put(7.0,18.0){\makebox(0,0)[c]{$\pi_1$}}
  \put(55.0,18.0){\makebox(0,0)[c]{$\pi_2$}}
  \put(25.0,25.0){\vector(-1,-1){18}}
  \put(37.0,25.0){\vector(1,-1){18}}
 \end{picture}
\end{equation}
with coordinates $(x^{AB},\lambda_A)$ on $\FC^6\times \CPP^3$ and $(z^A,\lambda_A)$ on $\CT^6$. The projection $\pi_2$ is again trivial and $\pi_1$ is given by
\begin{equation}
 \pi_1(x^{AB},\lambda_A)=(z^A,\lambda_A):=(x^{AB}\lambda_B,\lambda_A)~.
\end{equation}
The definition of $z^A$ implies the relation $z^A\lambda_A=0$, and therefore $\CT^6$ is a quadric in the total space of the rank 4 vector bundle $\FC^4\otimes \CO(1)\rightarrow \CPP^3$. The vector fields along the fibration $\pi_1$ are spanned by
\begin{equation}
 V^A=\lambda_B\dpar^{AB}~.
\end{equation}
The twistor space $\CT^6$ has been studied long ago by many authors and a complete list of references is found in Saemann \& Wolf\wpcite{Saemann:2011nb}, see also the discussion in Mason \wpetal\wpcite{Mason:2011nw}.

\subsection{Deriving a (2,0)-theory}

Let us now outline the construction of a $(2,0)$-theory, omitting technical details. We start from a topologically trivial holomorphic principal 2-bundle $\hat \CCP$ over $\CT^6$, which becomes holomorphically trivial when restricted to any $\CPP^3\embd \CT^6$. After pulling $\hat \CCP$ back along $\pi_2$, we can perform a gauge transformation rendering the \v Cech cocycles trivial, but creating a connection on $\pi_2^*\hat \CCP$ which consists of a globally defined 1-form $A$ and a globally defined 2-form $B$. These are flat on $\FC^6\times \CPP^3$ and contain in particular 1- and 2-form potentials on $\FC^6$, whose curvature 2-form satisfies the fake curvature condition and whose 3-form curvature satisfies the self-duality equation\wpcite{Saemann:2012uq}.

This construction is readily extend to the supersymmetric case by replacing the spaces in \eqref{eq:DoubleFibration2} by corresponding superspaces\wpcite{Saemann:2012uq}. The result on spacetime is precisely the field content of the $(2,0)$ tensor multiplet, and on superspace, one has the equations
\begin{equation}
 H=*H~,~~~\CF=0~,~~~\nablas\psi=0~,~~~\square \phi=0~.
\end{equation}
There is also a higher version of theorem~\ref{thm:Ward}:
\begin{theorem}
 Topologically trivial principal 2-bundles over $\CT^6$ which become holomorphically trivial when restricted to any $\CPP^3\embd \CT^6$ are in one-to-one correspondence with solutions to manifestly $\CN=(2,0)$ superconformal field equations on $\FC^6$, modulo isomorphisms on both sides.
\end{theorem}
Note that we presented only one direction of the proof of this theorem, and the inverse direction involves a higher Poincar\'e lemma\wpcite{Demessie:2014ewa} for relative categorified connections.

While the field equations for finite-dimensional strict Lie 2-groups are not yet very convincing, we have effectively reduced the search for a (2,0)-theory to a search for the appropriate higher gauge structure. That is, given any higher gauge structure, we construct the corresponding higher gauge theory as described in Section~\ref{sec:2} and then perform the Penrose--Ward transform by generalizing the discussion in Section~\ref{sec:3} to obtain corresponding (2,0)-theories.

\subsection{Further reading}

A detailed explanation of twistor space together with the Penrose--Ward transform in a language close to the one we used above is found in Popov \& Saemann\wpcite{Popov:2004rb} and in particular in Wolf\wpcite{Wolf:2010av}. The twistor space for self-dual 3-forms is discussed in detail in Saemann \& Wolf\wpcite{Saemann:2011nb} and Mason \wpetal\wpcite{Mason:2011nw}. The Penrose--Ward transform for various generalizations of the gauge structure is discussed in our papers\wpcite{Saemann:2013pca,Jurco:2014mva,Jurco:2016qwv}, with the last paper giving a very general account that subsumes all previous ones. Very useful general reference for twistor geometry and its application in field theory are the textbooks\wpcite{Penrose:1985jw,Penrose:1986ca,Ward:1990vs,Mason:1991rf}.

\section{Higher quantization}

Let us now turn to a slightly different topic, the quantization of multisymplectic manifolds. This also uses the language which we developed in the Section~\ref{sec:2} and its result should produce the appropriate gauge structure for M2- and M5-branes.

\subsection{Motivation: Fuzzy funnel in M-theory}

Let us return once more to the monopole configuration in type IIA superstring theory, in which $k$ D2-branes end on a D4-brane,
\begin{equation}
\begin{tabular}{rcccccccc}
& 0 & 1 & 2 & 3 & 4 & 5 & 6 & \ldots\\
D2 & $\times$ & & & & & $\times$ & $\times$ \\
D4 & $\times$ & $\times$ & $\times$ & $\times$ & & $\times$ &
\end{tabular}
\end{equation}
As discussed previously, the underlying dynamics are described from the perspective of the D2-brane by the Nahm equation, and after gauge fixing $A_s=0$, we have
\begin{equation}
 \dder{s}X^i=\tfrac{1}{2}\eps^{ijk}[X^j,X^k]~.
\end{equation}
The scalar fields $X^i$, taking values in $\au(k)$, describe the position of the $k$ D4-branes. In particular, if the $X^i$ can be diagonalized simultaneously, the $j$-th eigenvalue of $X^i$ is the position of the $j$th D2-brane in the $x^i$-direction. A solution to this equation is readily found by a factorization ansatz\wpcite{Myers:1999ps}:
\begin{equation}
 X^i(s)=\frac{1}{s}\,\tau^i \qquad \mbox{with} \qquad
\tau^i=\eps^{ijk}\,[\tau^j,\tau^k]~.
\end{equation}
This solutions suggests that the above picture of D2-branes ending perpendicularly on D4-branes is too naive and has to be modified as follows. The radial function indicates a funnel-like shape of the D2-branes opening up onto the D4-branes. Moreover, the $\tau^i$ form a representation of $\asu(2)$, and a more precise analysis suggests that this representation has to be irreducible. Such matrices form coordinates on a fuzzy sphere\footnote{A noncommutative version of the sphere $S^2\cong \CPP^1$ obtained \wpeg\ by geometric quantization with prequantum line bundle $\CO(k)$ as explained e.g.\ in our paper\wpcite{DeBellis:2010pf}.}. That is, each point of the worldvolume of the D2-brane polarizes into a fuzzy sphere, providing a transition between the two spatial dimensions of the D2-brane and the four spatial dimensions of the D4-brane.

We are now interested in the lift of this situation to M-theory. The configuration obtained by choosing the $x^4$ direction as the M-theory direction is 
\begin{equation}
\begin{tabular}{rccccccc}
${\rm M}$ & 0 & 1 & 2 & 3 & \phantom{(}4\phantom{)} & 5 & 6 \\
M2 & $\times$ & & & & & $\times$ & $\times$ \\
M5 & $\times$ & $\times$ & $\times$ & $\times$ & $\times$ & $\times$ 
\end{tabular}
\end{equation}
Recall the description from the M2-brane perspective suggested by Basu and Harvey:
\begin{equation}
 \dder{s}X^\mu=\tfrac{1}{3!}\eps^{\mu\nu\kappa\lambda}[X^\nu,X^\kappa,X^\lambda]~,
\end{equation}
$\mu=1,\ldots,4$. Its factorization solution is
\begin{equation}
 X^i(s)=\frac{1}{\sqrt{2s}} \,\tau^\mu \qquad \mbox{with} \qquad \tau^\mu=\eps^{\mu\nu\kappa\lambda}\,[\tau^\nu,\tau^\kappa,\tau^\lambda] \ .
\end{equation}
This suggests a similar interpretation as for the D2-D4-brane system: The M2-brane opens as a funnel onto the M5-brane, with each point of the worldvolume polarizing into a fuzzy 3-sphere. The problem with this interpretation is that no fully satisfactory quantization of the 3-sphere is known as of now. This would require a consistent approach to the quantization of multisymplectic manifolds and we turn to these in the following.

\subsection{Observables on 2-plectic manifolds}

A \wpem{multisymplectic manifold} $(M,\varpi)$ is a manifold endowed with a closed, non-degenerate differential form $\varpi$:
\begin{equation}
 \dd \varpi=0\eand \iota_X\varpi=0 \Leftrightarrow X=0~.
\end{equation}
If the form $\varpi$ is of degree $p+1$, we also call the multisymplectic manifold $(M,\varpi)$ \wpem{$p$-plectic}. In this nomenclature, symplectic manifolds are called 1-plectic manifolds. In the following, we shall focus on 2-plectic manifolds such as $\FR^3$ and $S^3$, for which the multisymplectic 3-form is simply the volume form.

As a first step, we should develop a notion of observables on such 2-plectic manifolds. This has been developed to various degrees\wpcite{Cantrijn:1996aa,Cantrijn:1999aa,Baez:2008bu}; more details are also found in Ritter \& Saemann\wpcite{Ritter:2015ffa,Ritter:2015ymv}.

Recall that if the phase space is a symplectic manifold $(M,\omega)$, the observables are given by the smooth functions $\CC^\infty(M)$. The symplectic form induces a Lie algebra structure on $\CC^\infty(M)$ as follows. To each observable $f\in\CC^\infty(M)$, we associate a corresponding Hamiltonian vector field $X_f$ such that $\iota_{X_f}\omega=\dd f$. The Lie bracket on $\CC^\infty(M)$ is then defined as 
\begin{equation}
 \{f,g\}=\iota_{X_f}\iota_{X_g}\omega~.
\end{equation}
This Lie bracket turns out to be compatible with the associative product on $\CC^\infty(M)$ and therefore induces a Poisson structure.

If we want to introduce an analogous structure on a 2-plectic manifold $(M,\varpi)$, we are naturally led to considering those 1-forms $\alpha$, which have a Hamiltonian vector field $X_\alpha$ such that $\iota_{X_\alpha}\varpi=\dd \alpha$. We denote the set of such Hamiltonian 1-forms by $\Omega^1_{\rm Ham}(M)$. With the help of the Hamiltonian vector fields, we can write down a 2-bracket:
\begin{equation}
  \mu_2:\Omega^1_{\rm Ham}(M)\wedge \Omega^1_{\rm Ham}(M)\rightarrow \Omega^1_{\rm Ham}(M)~,~~~\mu_2(\alpha,\beta)=\iota_{X_\alpha}\iota_{X_\beta}\varpi~.
\end{equation}
This bracket does not satisfy the Jacobi identity, but rather
\begin{equation}
 \mu_2(\mu_2(\alpha,\beta),\gamma)+\mu_2(\mu_2(\beta,\gamma),\alpha)+\mu_2(\mu_2(\gamma,\beta),\alpha)=\dd \iota_{X_\alpha}\iota_{X_\beta}\iota_{X_\gamma}\varpi~.
\end{equation}
This is reminiscent of a Lie 2-algebra, and indeed, on the complex
\begin{equation}
 \CC^\infty(M)\xrightarrow{~\dd~} \Omega^1_{\rm Ham}(M)~,
\end{equation}
we can introduce the non-trivial brackets
\begin{equation}
 \mu_1(f)=\dd f~,~~~\mu_2(\alpha,\beta)=\iota_{X_\alpha}\iota_{X_\beta}\varpi~,~~~
 \mu_3(\alpha,\beta,\gamma)=\iota_{X_\alpha}\iota_{X_\beta}\iota_{X_\gamma}\varpi~,
\end{equation}
which satisfy the higher Jacobi identities of a semistrict Lie 2-algebra. 

\begin{exercise}
 Verify the homotopy Jacobi relations \eqref{eq:homotopyJacobi} for the above Lie 2-algebra.
\end{exercise}

It is now natural to assume that this Lie 2-algebra takes over the role of the (categorified) Lie algebra of observables on a 2-plectic manifold. There are, in fact, many further observations that support this point of view.

An open problem in this context is the definition of an associative product on this Lie 2-algebra which is compatible with the Lie 2-algebra structure. It is, however, not even clear, whether it is reasonable to expect such a product. After all, the equations of classical mechanics only make explicit use of the Poisson bracket.

\subsection{Quantization of multisymplectic manifolds}

We now come to a quick review of what is known about higher geometric quantization of multisymplectic manifolds, and we go through the cases symplectic and 2-plectic manifolds in parallel.

\

\noindent\begin{tabularx}{\textwidth}{|X|X|}
 \hline
 Symplectic Geometry & 2-plectic Geometry\\
 \hline
 Symplectic manifold $(M,\omega)$ which satisfies the quantization condition $\omega\in H^2(M,\RZ)$ & 2-plectic manifold $(M,\varpi)$ which satisfies the quantization condition $\varpi\in H^3(M,\RZ)$ \\
 \hline
 Prequantum line bundle $(L,\nabla)$ with first Chern class $\omega$: $F=\nabla^2=2\pi\di\omega$ &
 Prequantum line bundle gerbe $(\CCL,B)$ with Dixmier--Douady class $\varpi$: $H=\dd B=2\pi\di\varpi$\\
 \hline
 Pre-Hilbert space is the set of sections of this prequantum line bundle $\CH=\Gamma(L)$, which can be regarded as morphisms from the trivial line bundle to $L$.&
 Pre-Hilbert space is the set of sections of the prequantum line bundle gerbe. These sections are identified with morphisms from the trivial line bundle gerbe to $\CCL$, which in turn are bundle gerbe modules or twisted vector bundles.\\
 \hline 
 The observables are (real) endomorphisms on the prequantum line bundle, and given by sections of the trivial line bundle, or, morphisms from the trivial line bundle to itself. The resulting set is $\CC^\infty(M)$. &
 Correspondingly, observables should be sections of the trivial line bundle gerbe. Real such sections can be shown to contain the expected classical observables
 $\CC^\infty(M)\oplus\Omega^1(M)$. \\
 \hline
 Lie algebra structure $\{-,-\}$ on $\CC^\infty(M)$&
 Lie 2-algebra structure $\mu_1$, $\mu_2$, $\mu_3$ on $\CC^\infty(M)\oplus\Omega^1(M)$\\
 \hline
 square integrable sections & unknown, possibly direct square integrable sections\\
 \hline
 For quantum mechanics, the pre-Hilbert space is too big and needs to be reduced to half its size. This is done via a polarization, and in the case of Kähler polarization, we reduce $\CH$ to holomorphic sections. & The notion of polarization is mostly unclear for 2-plectic manifold. An answer can possibly be found, however, when working with categorified spaces\wpcite{Ritter:2015ffa}.\\
 \hline
 coherent states $|z\rangle$ &
 unknown\\
 \hline
 quantization relation, \wpeg\ $f=\tr(\frac{|z\rangle\langle z|}{\langle z|z\rangle}\hat{f})$ &unknown
 \\
 \hline
\end{tabularx}

\

There are a couple of further issues in this picture. If we want to construct the Hilbert space of a multisymplectic manifold for a 3-form which is not torsion, then the corresponding line bundle gerbe has infinite-dimensional bundle gerbe modules as sections. In these cases, things are very hard to get under control, both abstractly and for explicit computations\wpcite{Bunk:2016rta}.

However, something that we can already learn from this picture is that the symmetry group covering the isometries on $\FR^3$ acting on the prequantum 2-Hilbert space of quantized $\FR^3$ is a string 2-group model of $\sSpin(3)$\wpcite{Bunk:2016rta}. Comparing with the analogue statements for D-branes, this suggests that the correct higher gauge group for M-brane models is a string 2-group model.

\subsection{Transgression to loop space}

One potential solution to the problems encountered in higher quantization is to apply a trick, called ``transgression,'' which allows to discuss higher quantization as ordinary quantization on loop space.

This transgression is a map that transfers de Rham cohomology on a manifold $M$ to the manifold's loop space $\CL M=\CC^\infty(S^1,M)$. We start from the double fibration
\begin{equation}
 \begin{picture}(50,40)
  \put(0.0,0.0){\makebox(0,0)[c]{$M$}}
  \put(64.0,0.0){\makebox(0,0)[c]{$M$}}
  \put(34.0,33.0){\makebox(0,0)[c]{$\CL M \times S^1$}}
  \put(7.0,18.0){\makebox(0,0)[c]{${\rm ev}$}}
  \put(55.0,18.0){\makebox(0,0)[c]{${\rm pr}$}}
  \put(25.0,25.0){\vector(-1,-1){18}}
  \put(37.0,25.0){\vector(1,-1){18}}
 \end{picture}
\end{equation}
with the trivial projection ${\rm pr}$ and the evaluation map ${\rm ev}:\CL M\times S^1\rightarrow M$. Transgression now maps a $p+1$-form $\alpha$ on $M$ to a $p$-form on loop space $\CL M$ according to the formula
\begin{equation}
 \CT \alpha:=\oint_{S^1}{\rm ev}^*\alpha~.
\end{equation}
A more explicit description is the following. Note that there is a natural tangent vector $\xd\in \CL TM\cong T\CL M$ to every point $x:S^1\embd M$ in loop space. Correspondingly, 
\begin{equation}
 (\CT \alpha)(X^1,\ldots, X^p):=\oint_{S^1}\dd \tau~\alpha(x(\tau))\big(X^1(\tau),\ldots,X^p(\tau),\xd(\tau)\big)~.
\end{equation}
The transgression map is in fact a chain map: $\delta\circ \CT=\CT\circ \dd$, where $\dd$ and $\delta$ are the de Rham differentials on $M$ and $\CL M$, respectively. 

If we are merely interested in the loops themselves, instead of their parametrization, we can factor out reparametrization transformations to obtain knot space 
\begin{equation}
 \CK M:= \CL M/{\mathsf{Diff}^+(S^1)}~.
\end{equation}
For details on this and the rigorous definitions, see Brylinski\wpcite{0817647309}. Fortunately, the images of the transgression map are invariant under reparametrization transformations and therefore descend to loop space. To work on knot space, we can simply use loop space expressions, making sure that all quantities are reparametrization invariant.

The idea is now to consider the symplectic manifold $(\CK M,\CT \varpi)$ instead of the 2-plectic manifold $(M,\varpi)$, and perform geometric quantization as usual.

As a first step, we should consider the observables, which will be $\CC^\infty(\CL M)$. This vector space receives a Lie algebra structure by the usual construction of the Poisson bracket. Note that $\CT \varpi$ is degenerate on loop space, as any vector field of the form $X=\oint \alpha(\tau)\xd^i(\tau)\frac{\delta}{\delta x^i(\tau)}$ in some local coordinates $x^i(\tau)$ lies in the kernel of $\CT\omega:T(\CL M)\rightarrow T^*(\CL M)$. However, these vector fields generate reparametrizations and after restricting to knot space, $\CT\varpi$ is non-degenerate. Its inverse defines a Poisson bivector and the resulting Poisson bracket is compatible with the Lie 2-algebra introduced above in the sense that
\begin{equation}
 \{\CT\alpha,\CT\beta\}_{\CT\varpi}=\CT(\mu_2(\alpha,\beta))~,
\end{equation}
where $\mu_2$ is the Lie 2-algebra product on 1-forms induced by the 2-plectic form $\varpi$.

\subsection{Towards a quantization of loop space}

Note that based loop spaces of Lie groups have been quantized before in the mathematical literature. There are two differences to our situation. First, we are working with knot space instead of the based loop space. Second, there is a natural symplectic structure on loop space,
\begin{equation}
 \omega=\oint\dd \tau~g_{ij}(x(\tau)) \delta x^i(\tau)\wedge \delta \xd^j(\tau)~,
\end{equation}
in some local Cartesian coordinates $x^i(\tau)$, where $g_{ij}(x)$ is a metric on the underlying manifold. We shall be working with the transgressed 2-plectic form instead. (In particular cases, e.g.\ when the manifold is a simple Lie group, this form agrees with the natural symplectic form on loop space\wpcite{Pressley:1988aa}.)

As mentioned above, we have to reduce the prequantum Hilbert space to a true Hilbert space by introducing a polarization. This can be done by introducing a complex structure on the symplectic manifold, which allows us to restrict the general smooth sections of the prequantum line bundle to holomorphic sections. The complex structure has to be compatible with the symplectic structure and therefore, the manifolds we quantize are Kähler manifolds.

While there is no obvious candidate for a complex structure on loop space, there is one on the knot space of 3-dimensional manifolds. The tangent bundle $T\CK M$ is at each point of each knot spanned by a 2-dimensional plane perpendicular to the tangent vector to the knot. We can thus define a complex structure, which rotates the vectors in the plane at each point of each knot by $\frac{\pi}{2}$. This operation is consistent and squares to $-\id$. Moreover, together with the transgressed 2-plectic form, this yields indeed a Kähler structure on knot space $\CK M$. See again Brylinski\wpcite{0817647309} for a detailed discussion of this point.

In principle, we can now proceed and try to define the vector space underlying a Hilbert space. A definition of an inner product will be more subtle since we do not have a reparametrization invariant measure on loop or knot space.

To be more concrete, let us focus on the example of $\FR^3$ with 2-plectic form the volume form, $\varpi=\tfrac{1}{3!}\eps_{ijk}\dd x^i\wedge \dd x^j\wedge \dd x^k$ in standard cartesian coordinates. We transgress this to the following symplectic form on knot space:
\begin{equation}
 \CT\varpi=\oint \dd \tau~\tfrac{1}{2} \eps_{ijk} \delta x^i(\tau)\wedge \delta x^j(\tau) \xd^k(\tau)~.
\end{equation}
The corresponding inverse bivector induces the Poisson bracket
\begin{equation}\label{eq:loop_space_poisson}
 \{x^i(\tau),x^j(\sigma)\}=\eps^{ijk}\frac{\xd^k(\tau)}{|\xd(\tau)|}\delta(\tau-\sigma)
\end{equation}
on $\CC^\infty(\CK M)$. 

Recall that a quantization map is a Lie algebra homomorphism to first order in $\hbar$ between the Poisson algebra of classical observables and the Lie algebra of quantum observables. Moreover, on coordinate functions it is usually an exact Lie algebra homomorphism. We therefore expect that
\begin{equation}
 [\hat x^i(\tau),\hat x^j(\sigma)]=-i\hbar\eps^{ijk}\frac{\hat\xd^k(\tau)}{|\hat \xd(\tau)|}\delta(\tau-\sigma)~.
\end{equation}
This agrees with various computations in M-theory\wpcite{Bergshoeff:2000jn,Kawamoto:2000zt,Matsuo:0010040}.

It is now unfortunately a rather difficult problem to construct holomorphic sections of the trivial line bundle over $\CK\FR^3$. Interestingly, such functions can be constructed using twistor spaces, see\wpcite{Saemann:2012ab}.

As a final consistency check, let us discuss the reduction of the M-brane picture to string theory. That is, we compactify one direction of $\FR^3$, say $x^3$, on a circle and force all knots to be oriented in this direction:
\begin{equation}
 x^i(\tau)=x^i_0+2\pi R\tau\delta^{i3}~,
\end{equation}
where $R$ is the radius of the loop. If we plug these restricted knots into the knot space Poisson bracket \eqref{eq:loop_space_poisson} and restrict to zero modes by integrating over the loop parameter, we obtain
\begin{equation}
 \oint \dd \tau\oint \dd \sigma \{x^a(\tau),x^b(\sigma)\}=4\pi^2 R^2\{x^a_0,x^b_0\}=\oint \dd \tau\eps^{ab3}2\pi R \tau=4\pi^2R^2\eps^{ab}~.
\end{equation}
for $a,b\in \{1,2\}$. That is,
\begin{equation}
 \{x^a_0,x^b_0\}=\eps^{ab}~,
\end{equation}
and we recovered the Poisson algebra on $\FR^2$.

\subsection{Further reading}

The 2-vector spaces formed by sections of prequantum bundle gerbes, which should underlie categorified Hilbert spaces, were first developed in detail by Waldorf\wpcite{Waldorf:2007aa} and then technically developed further to prequantum 2-Hilbert spaces\wpcite{Bunk:2016rta}; see also the papers\wpcite{Fiorenza:2013kqa,Fiorenza:1304.6292} for a detailed account of higher prequantization.

The loop space approach to quantization as sketched above was studied in Saemann \& Szabo\wpcite{Saemann:2012ab,Saemann:2012ex}. 

There is also a generalized notion of Poisson bracket, known as \wpem{Nambu--Poisson bracket}, and the problems one faces trying to quantize it properly are summarized in our paper\wpcite{DeBellis:2010pf}. The overlap with multisymplectic geometry is only partial, as explained \wpeg\ in Ritter \& Saemann\wpcite{Ritter:2015ymv}. 

\appendices

\subsection{Higher Lie algebras and 3-Lie algebras}\label{app:A}

During both the workshops at the Erwin Schrödinger Institute and at Tohoku University, several participants asked if there was a relationship between the 3-Lie algebras of the M2-brane models\wpcite{Bagger:2007jr,Gustavsson:2007vu} and the categorified Lie $n$-algebras discussed in these lectures. Let us therefore summarize the relevant statements.

Recall that a \wpem{3-Lie algebra}\wpcite{Filippov:1985aa} is a vector space $\CA$ endowed with a ternary, totally antisymmetric bracket $[-,-,-]:\CA^{\wedge 3}\rightarrow \CA$. This bracket satisfies the fundamental identity
\begin{equation}
 [a,b,[c,d,e]]=[[a,b,c],d,e]+[c,[a,b,d],e]+[c,d,[a,b,e]]
\end{equation}
for all $a,b,c,d,e\in \CA$, which implies that the inner derivations $D(a,b)$, which act on $c\in \CA$ according to
\begin{equation}
 D(a,b)c=[a,b,c]
\end{equation}
form a Lie algebra $\frg_\CA$. We can further equip $\CA$ with a metric $(-,-)$ satisfying
\begin{equation}
 ([a,b,c],d)+(c,[a,b,d])=0~.
\end{equation}

In\wpcite{Cherkis:2008qr}, a generalization was defined, in which the 3-bracket is only antisymmetric in its first two slots. It was noted in\wpcite{deMedeiros:2008zh} that the resulting generalized metric 3-Lie algebras are in one-to-one correspondence with metric Lie algebras $\frg$ and faithful orthogonal $\frg$-modules.

This observation was then extended in\wpcite{Palmer:2012ya} to the statement that each generalized metric 3-Lie algebra has an underlying metric strict Lie 2-algebra $\CA\xrightarrow{~\dpar~} \frg_\CA$ with metrics on $\CA$ and $\frg_\CA$ and non-trivial higher products
\begin{equation}
 \mu_2(D(a,b),D(c,d))=[D(a,b),D(c,d)]\eand \mu_2(D(a,b),c)=[a,b,c]~.
\end{equation}

Inversely, on each strict Lie 2-algebra $\frh\rightarrow \frg$ with metrics $\lbr-,-\rbr$ and $(-,-)$ on $\frg$ and $\frh$, respectively, there is a bilinear map $D:\frh\wedge \frh\rightarrow \frg$ such that
\begin{equation}
 \lbr g_1,D(a,b)\rbr=-(\mu_2(g_1,a), b)~.
\end{equation}
A corresponding 3-bracket is then defined as
\begin{equation}
 [a,b,c]:=\mu_2(D(a,b),c)~.
\end{equation}

Altogether, metric 3-Lie algebras and their generalizations are strict metric Lie 2-algebras, and the nomenclature is rather unfortunate. Moreover, the 3-bracket on a 3-Lie algebra is not related to the higher product $\mu_3$, which vanishes for strict Lie 2-algebras.

There is, however, an interesting class of examples of 3-Lie algebras, in which both ternary maps can be made to agree. Consider the 3-Lie algebra defined originally in\wpcite{Awata:1999dz}, where $\CA=\agl(N,\FC)$ and 
\begin{equation}
 [a,b,c]:=\tr(a)[b,c]+\tr(b)[c,a]+\tr(c)[a,b]~.
\end{equation}
As observed in\wpcite{Ritter:2013wpa}, this 3-Lie algebra can actually be extended to a semistrict Lie 2-algebra on the complex $\agl(N,\FC)\xrightarrow{~\id~}\agl(N,\FC)$ with higher products
\begin{equation}
 \begin{aligned}
  \mu_1(v)&=v~,\\
  \mu_2(w_1,w_2)&=\tr(w_1)w_2-\tr(w_2)w_1+[w_1,w_2]~,\\
  \mu_2(v,w)&=-(\tr(v)w-\tr(w)v+[v,w])~,\\
  \mu_3(w_1,w_2,w_3)&=\tr(w_1)[w_2,w_3]+\tr(w_2)[w_3,w_1]+\tr(w_3)[w_1,w_2]~,
 \end{aligned}
\end{equation}
where we denoted elements from the left and the right vector space in $\agl(N,\FC)\xrightarrow{~\id~}\agl(N,\FC)$ by $v_{1,2,3}$ and $w_{1,2,3}$, respectively. 

\bibliography{bigone}

\begin{thebibliography}{10}

\bibitem{Giveon:1998sr}
A.~Giveon and D.~Kutasov,
{\em Brane dynamics and gauge theory,}
\href{http://dx.doi.org/10.1103/RevModPhys.71.983}{Rev. Mod. Phys. {\bf 71}
  (1999) 983} [{\tt
  \href{http://www.arxiv.org/abs/hep-th/9802067}{hep-th/9802067}}].
%%CITATION = HEP-TH/9802067;%%

\bibitem{Berman:2007bv}
D.~S.~Berman,
{\em M-theory branes and their interactions,}
\href{http://dx.doi.org/10.1016/j.physrep.2007.10.002}{Phys. Rept. {\bf 456}
  (2008)~89} [{\tt \href{http://www.arxiv.org/abs/0710.1707}{0710.1707
  [hep-th]}}].
%%CITATION = 0710.1707;%%

\bibitem{Witten:1995zh}
E.~Witten,
{\em {Some comments on string dynamics},}
proceedings of ``Strings ‘95'', USC, 1995
[{\tt \href{http://www.arxiv.org/abs/hep-th/9507121}{hep-th/9507121}}].
%%CITATION = HEP-TH/9507121;%%

\bibitem{Strominger:1995ac}
A.~Strominger,
{\em Open p-branes,}
\href{http://dx.doi.org/10.1016/0370-2693(96)00712-5}{Phys. Lett. B {\bf 383}
  (1996)~44} [{\tt
  \href{http://www.arxiv.org/abs/hep-th/9512059}{hep-th/9512059}}].
%%CITATION = HEP-TH/9512059;%%

\bibitem{Witten:1995em}
E.~Witten,
{\em {Five-branes and M-theory on an orbifold},}
\href{http://dx.doi.org/10.1016/0550-3213(96)00032-6}{Nucl. Phys. B {\bf 463}
  (1996) 383} [{\tt
  \href{http://www.arxiv.org/abs/hep-th/9512219}{hep-th/9512219}}].
%%CITATION = HEP-TH/9512219;%%

\bibitem{Gawedzki:1987ak}
K.~Gawedzki,
{\em {Topological actions in two-dimensional quantum field theories},}
Nonperturbative quantum field theory (Carg\`ese, 1987), 101--141, NATO Adv.
  Sci. Inst. Ser. B Phys., 185, Plenum, New York, 1988.

\bibitem{Freed:1999vc}
D.~S.~Freed and E.~Witten,
{\em Anomalies in string theory with D-branes,}
Asian J. Math {\bf 3} (1999) 819 [{\tt
  \href{http://www.arxiv.org/abs/hep-th/9907189}{hep-th/9907189}}].
%%CITATION = HEP-TH/9907189;%%

\bibitem{Hilton:1962:227-255}
B.~Eckmann and P.~J.~Hilton,
{\em Group-like structures in general categories I: Multiplications and
  comultiplications,}
\href{http://dx.doi.org/10.1007/BF01451367}{Math. Ann. {\bf 145} (1962) 227}.

\bibitem{Mackaay:2000ac}
M.~Mackaay and R.~Picken,
{\em {The holonomy of gerbes with connections},}
\href{http://dx.doi.org/10.1006/aima.2002.2085}{Adv. Math. {\bf 170} (2002)
  287} [{\tt
  \href{http://www.arxiv.org/abs/math.DG/0007053}{math.DG/0007053}}].
%%CITATION = MATH/0007053;%%

\bibitem{Bagger:2007jr}
J.~Bagger and N.~D.~Lambert,
{\em Gauge symmetry and supersymmetry of multiple M2-branes,}
\href{http://dx.doi.org/10.1103/PhysRevD.77.065008}{Phys. Rev. D {\bf 77}
  (2008) 065008} [{\tt \href{http://www.arxiv.org/abs/0711.0955}{0711.0955
  [hep-th]}}].
%%CITATION = 0711.0955;%%

\bibitem{Gustavsson:2007vu}
A.~Gustavsson,
{\em Algebraic structures on parallel M2-branes,}
\href{http://dx.doi.org/10.1016/j.nuclphysb.2008.11.014}{Nucl. Phys. B {\bf
  811} (2009)~66} [{\tt \href{http://www.arxiv.org/abs/0709.1260}{0709.1260
  [hep-th]}}].
%%CITATION = 0709.1260;%%

\bibitem{Aharony:2008ug}
O.~Aharony, O.~Bergman, D.~L.~Jafferis, and J.~M.~Maldacena,
{\em {$\CN=6$ superconformal Chern-Simons-matter theories, M2-branes and their
  gravity duals},}
\href{http://dx.doi.org/10.1088/1126-6708/2008/10/091}{JHEP {\bf 0810} (2008)
  091} [{\tt \href{http://www.arxiv.org/abs/0806.1218}{0806.1218 [hep-th]}}].
%%CITATION = 0806.1218;%%

\bibitem{Howe:1997ue}
P.~S.~Howe, N.~D.~Lambert, and P.~C.~West,
{\em The self-dual string soliton,}
\href{http://dx.doi.org/10.1016/S0550-3213(97)00750-5}{Nucl. Phys. B {\bf 515}
  (1998) 203} [{\tt
  \href{http://www.arxiv.org/abs/hep-th/9709014}{hep-th/9709014}}].
%%CITATION = HEP-TH/9709014;%%

\bibitem{Basu:2004ed}
A.~Basu and J.~A.~Harvey,
{\em The M2-M5 brane system and a generalized Nahm's equation,}
\href{http://dx.doi.org/10.1016/j.nuclphysb.2005.02.007}{Nucl. Phys. B {\bf
  713} (2005) 136} [{\tt
  \href{http://www.arxiv.org/abs/hep-th/0412310}{hep-th/0412310}}].
%%CITATION = HEP-TH/0412310;%%

\bibitem{JSTOR:1998201}
M.~Batchelor,
{\em The structure of supermanifolds,}
\href{http://dx.doi.org/10.2307/1998201}{Trans. Am. Math. Soc. {\bf 253} (1979)
  329}.

\bibitem{Bonavolonta:2012fh}
G.~Bonavolonta and N.~Poncin,
{\em {On the category of Lie n-algebroids},}
\href{http://dx.doi.org/10.1016/j.geomphys.2013.05.004}{J Geom. Phys. {\bf 73}
  (2013) 70–90} [{\tt \href{http://www.arxiv.org/abs/1207.3590}{1207.3590
  [math.DG]}}].
%%CITATION = ARXIV:1207.3590;%%

\bibitem{Baez:2003aa}
J.~Baez and A.~S.~Crans,
{\em Higher-dimensional algebra VI: Lie 2-algebras,}
\href{http://tac.mta.ca/tac/volumes/12/15/12-15.pdf}{Th. App. Cat. {\bf 12}
  (2004) 492} [{\tt
  \href{http://www.arxiv.org/abs/math.QA/0307263}{math.QA/0307263}}].

\bibitem{Baez:2004in}
J.~C.~Baez and U.~Schreiber,
{\em Higher gauge theory: 2-connections on 2-bundles,}
{\tt \href{http://www.arxiv.org/abs/hep-th/0412325}{hep-th/0412325}}.
%%CITATION = HEP-TH/0412325;%%

\bibitem{Baez:2010ya}
J.~C.~Baez and J.~Huerta,
{\em {An invitation to higher gauge theory},}
\href{http://dx.doi.org/10.1007/s10714-010-1070-9}{Gen. Relativ. Gravit. {\bf
  43} (2011) 2335} [{\tt \href{http://www.arxiv.org/abs/1003.4485}{1003.4485
  [hep-th]}}].
%%CITATION = ARXIV:1003.4485;%%

\bibitem{Saemann:2010cp}
C.~Saemann,
{\em {Constructing self-dual strings},}
\href{http://dx.doi.org/10.1007/s00220-011-1257-2}{Commun. Math. Phys. {\bf
  305} (2011) 513} [{\tt \href{http://www.arxiv.org/abs/1007.3301}{1007.3301
  [hep-th]}}].
%%CITATION = ARXIV:1007.3301;%%

\bibitem{Roytenberg:0203110}
D.~Roytenberg,
{\em On the structure of graded symplectic supermanifolds and Courant
  algebroids,}
in: ``Quantization, Poisson Brackets and Beyond,'' ed.\ Theodore Voronov,
  Contemp. Math., Vol. 315, Amer. Math. Soc., Providence, RI, 2002
[{\tt \href{http://www.arxiv.org/abs/math.SG/0203110}{math.SG/0203110}}].

\bibitem{Lada:1992wc}
T.~Lada and J.~Stasheff,
{\em {Introduction to sh Lie algebras for physicists},}
\href{http://dx.doi.org/10.1007/BF00671791}{Int. J. Theor. Phys. {\bf 32}
  (1993) 1087} [{\tt
  \href{http://www.arxiv.org/abs/hep-th/9209099}{hep-th/9209099}}].
%%CITATION = HEP-TH/9209099;%%

\bibitem{Lada:1994mn}
T.~Lada and M.~Markl,
{\em {Strongly homotopy Lie algebras},}
\href{http://dx.doi.org/10.1080/00927879508825335}{Commun. Alg. {\bf 23} (1995)
  2147} [{\tt \href{http://www.arxiv.org/abs/hep-th/9406095}{hep-th/9406095}}].
%%CITATION = HEP-TH/9406095;%%

\bibitem{Zwiebach:1992ie}
B.~Zwiebach,
{\em {Closed string field theory: Quantum action and the B-V master equation},}
\href{http://dx.doi.org/10.1016/0550-3213(93)90388-6}{Nucl. Phys. B {\bf 390}
  (1993)~33} [{\tt
  \href{http://www.arxiv.org/abs/hep-th/9206084}{hep-th/9206084}}].
%%CITATION = HEP-TH/9206084;%%

\bibitem{Jurco:2014mva}
B.~Jurco, C.~Saemann, and M.~Wolf,
{\em {Semistrict higher gauge theory},}
\href{http://dx.doi.org/10.1007/JHEP04(2015)087}{JHEP {\bf 1504} (2015) 087}
  [{\tt \href{http://www.arxiv.org/abs/1403.7185}{1403.7185 [hep-th]}}].
%%CITATION = ARXIV:1403.7185;%%

\bibitem{Bojowald:0406445}
M.~Bojowald, A.~Kotov, and T.~Strobl,
{\em Lie algebroid morphisms, Poisson sigma models, and off-shell closed gauge
  symmetries,}
\href{http://dx.doi.org/10.1016/j.geomphys.2004.11.002}{J. Geom. Phys. {\bf 54}
  (2005) 400} [{\tt
  \href{http://www.arxiv.org/abs/math.DG/0406445}{math.DG/0406445}}].

\bibitem{Sati:0801.3480}
H.~Sati, U.~Schreiber, and J.~Stasheff,
{\em $L_\infty$-algebra connections and applications to String- and
  Chern-Simons $n$-transport,}
in: ``Quantum Field Theory,'' eds. B. Fauser, J. Tolksdorf and E. Zeidler, p.
  303, Birkh{\"a}user 2009
[{\tt \href{http://www.arxiv.org/abs/0801.3480}{0801.3480 [math.DG]}}].

\bibitem{Kotov:2010wr}
A.~Kotov and T.~Strobl,
{\em {Generalizing geometry - Algebroids and sigma models},}
in ``Handbook on Pseudo-Riemannian Geometry and Supersymmetry,'' ed. V. Cortes
[{\tt \href{http://www.arxiv.org/abs/1004.0632}{1004.0632 [hep-th]}}].
%%CITATION = ARXIV:1004.0632;%%

\bibitem{Gruetzmann:2014ica}
M.~Gruetzmann and T.~Strobl,
{\em {General Yang-Mills type gauge theories for p-form gauge fields: From
  physics-based ideas to a mathematical framework OR From Bianchi identities to
  twisted Courant algebroids},}
\href{http://dx.doi.org/10.1142/S0219887815500097}{Int. J. Geom. Meth. Mod.
  Phys. {\bf 12} (2014) 1550009} [{\tt
  \href{http://www.arxiv.org/abs/1407.6759}{1407.6759 [hep-th]}}].
%%CITATION = ARXIV:1407.6759;%%

\bibitem{Atiyah:1957}
M.~F.~Atiyah,
{\em Complex analytic connections in fibre bundles,}
\href{http://dx.doi.org/10.1090/S0002-9947-1957-0086359-5}{Trans. Amer. Math.
  Soc. {\bf 85} (1957) 181}.

\bibitem{Ritter:2015zur}
P.~Ritter, C.~Saemann, and L.~Schmidt,
{\em {Generalized higher gauge theory},}
\href{http://dx.doi.org/10.1007/JHEP04(2016)032}{JHEP {\bf 04} (2016) 032}
  [{\tt \href{http://www.arxiv.org/abs/1512.07554}{1512.07554 [hep-th]}}].
%%CITATION = ARXIV:1512.07554;%%

\bibitem{Fiorenza:2013nha}
D.~Fiorenza, H.~Sati, and U.~Schreiber,
{\em {Super Lie $n$-algebra extensions, higher WZW models, and super $p$-branes
  with tensor multiplet fields},}
\href{http://dx.doi.org/10.1142/S0219887815500188}{Int. J. Geom. Meth. Mod.
  Phys. {\bf 12} (2014) 1550018} [{\tt
  \href{http://www.arxiv.org/abs/1308.5264}{1308.5264 [hep-th]}}].
%%CITATION = ARXIV:1308.5264;%%

\bibitem{Baez:0307200}
J.~C.~Baez and A.~D.~Lauda,
{\em Higher-dimensional algebra V: 2-groups,}
\href{http://www.kurims.kyoto-u.ac.jp/EMIS/journals/TAC/volumes/12/14/12-14.pdf}{Th.
  App. Cat. {\bf 12} (2004) 423} [{\tt
  \href{http://www.arxiv.org/abs/math.QA/0307200}{math.QA/0307200}}].

\bibitem{Segal1968}
G.~Segal,
{\em Classifying spaces and spectral sequences,}
\href{http://eudml.org/doc/103878}{Pub. Math. IH{\'E}S {\bf 34} (1968) 105}.

\bibitem{Henriques:2006aa}
A.~Henriques,
{\em Integrating $L_\infty$-algebras,}
\href{http://dx.doi.org/10.1112/S0010437X07003405}{Comp. Math. {\bf 144} (2008)
  1017} [{\tt
  \href{http://www.arxiv.org/abs/math.CT/0603563}{math.CT/0603563}}].

\bibitem{Severa:2006aa}
P.~Severa,
{\em $L_\infty$-algebras as 1-jets of simplicial manifolds (and a bit beyond),}
{\tt \href{http://www.arxiv.org/abs/math.DG/0612349}{math.DG/0612349}}.

\bibitem{Jurco:2016qwv}
B.~Jurco, C.~Saemann, and M.~Wolf,
{\em {Higher groupoid bundles, higher spaces, and self-dual tensor field
  equations},}
\href{http://dx.doi.org/10.1002/prop.201600031}{Fortschr. Phys. {\bf 64} (2016)
  674} [{\tt \href{http://www.arxiv.org/abs/1604.01639}{1604.01639 [hep-th]}}].
%%CITATION = ARXIV:1604.01639;%%

\bibitem{Breen:math0106083}
L.~Breen and W.~Messing,
{\em Differential geometry of gerbes,}
\href{http://dx.doi.org/10.1016/j.aim.2005.06.014}{Adv. Math. {\bf 198} (2005)
  732} [{\tt
  \href{http://www.arxiv.org/abs/math.AG/0106083}{math.AG/0106083}}].

\bibitem{Baez:2002jn}
J.~C.~Baez,
{\em {Higher Yang-Mills theory},}
{\tt \href{http://www.arxiv.org/abs/hep-th/0206130}{hep-th/0206130}}.
%%CITATION = HEP-TH/0206130;%%

\bibitem{Aschieri:2003mw}
P.~Aschieri, L.~Cantini, and B.~Jur\v{c}o,
{\em Nonabelian bundle gerbes, their differential geometry and gauge theory,}
\href{http://dx.doi.org/10.1007/s00220-004-1220-6}{Commun. Math. Phys. {\bf
  254} (2005) 367} [{\tt
  \href{http://www.arxiv.org/abs/hep-th/0312154}{hep-th/0312154}}].
%%CITATION = HEP-TH/0312154;%%

\bibitem{Aschieri:2004yz}
P.~Aschieri and B.~Jurco,
{\em Gerbes, M5-brane anomalies and E(8) gauge theory,}
\href{http://dx.doi.org/10.1088/1126-6708/2004/10/068}{JHEP {\bf 10} (2004)
  068} [{\tt \href{http://www.arxiv.org/abs/hep-th/0409200}{hep-th/0409200}}].
%%CITATION = HEP-TH/0409200;%%

\bibitem{Bartels:2004aa}
T.~Bartels,
{\em Higher gauge theory I: 2-Bundles,} PhD thesis, University of
  California-Riverside (2006)
[{\tt \href{http://www.arxiv.org/abs/math.CT/0410328}{math.CT/0410328}}].

\bibitem{Jurco:2005qj}
B.~Jur\v{c}o,
{\em Crossed module bundle gerbes; Classification, string group and
  differential geometry,}
\href{http://dx.doi.org/10.1142/S0219887811005555}{Int. J. Geom. Meth. Mod.
  Phys. {\bf 08} (2011) 1079} [{\tt
  \href{http://www.arxiv.org/abs/math.DG/0510078}{math.DG/0510078}}].

\bibitem{Schreiber:2008aa}
U.~Schreiber and K.~Waldorf,
{\em Connections on non-abelian gerbes and their holonomy,}
\href{http://www.tac.mta.ca/tac/volumes/28/17/28-17.pdf}{Th. Appl. Cat. {\bf
  28} (2013) 476} [{\tt \href{http://www.arxiv.org/abs/0808.1923}{0808.1923
  [math.DG]}}].

\bibitem{Giraud:1971}
J.~Giraud,
{\em Cohomologie non ab\'elienne,} Springer, Berlin, 1971.

\bibitem{0817647309}
J.-L.~Brylinski,
{\em Loop spaces, characteristic classes and geometric quantization,}
  Birkh{\"a}user, Boston, 2007.

\bibitem{Murray:9407015}
M.~K.~Murray,
{\em Bundle gerbes,}
\href{http://dx.doi.org/10.1112/jlms/54.2.403}{J. Lond. Math. Soc. {\bf 54}
  (1996) 403} [{\tt
  \href{http://www.arxiv.org/abs/dg-ga/9407015}{dg-ga/9407015}}].

\bibitem{Murray:2007ps}
M.~K.~Murray,
{\em {An Introduction to bundle gerbes},}
in: {\em The many facets of geometry: A tribute to Nigel Hitchin,} eds.\ O.\
  Garcia-Prada, J.-P.\ Bourguignon and S.\ Salamon, Oxford University Press,
  Oxford, 2010
[{\tt \href{http://www.arxiv.org/abs/0712.1651}{0712.1651 [math.DG]}}].

\bibitem{Baez:0511710}
J.~C.~Baez and U.~Schreiber,
{\em Higher gauge theory,}
\href{http://dx.doi.org/10.1090/conm/431/08264}{Contemp. Math. {\bf 431}
  (2007)~7} [{\tt
  \href{http://www.arxiv.org/abs/math.DG/0511710}{math.DG/0511710}}].

\bibitem{Demessie:2016ieh}
G.~A.~Demessie and C.~Saemann,
{\em {Higher gauge theory with string 2-groups},}
{\tt \href{http://www.arxiv.org/abs/1602.03441}{1602.03441 [math-ph]}}.
%%CITATION = ARXIV:1602.03441;%%

\bibitem{Schreiber:2013pra}
U.~Schreiber,
{\em {Differential cohomology in a cohesive infinity-topos},}
Habilitation Thesis, 2011
[{\tt \href{http://www.arxiv.org/abs/1310.7930}{1310.7930 [math-ph]}}].

\bibitem{Saemann:2012uq}
C.~Saemann and M.~Wolf,
{\em {Non-abelian tensor multiplet equations from twistor space},}
\href{http://dx.doi.org/10.1007/s00220-014-2022-0}{Commun. Math. Phys. {\bf
  328} (2014) 527} [{\tt \href{http://www.arxiv.org/abs/1205.3108}{1205.3108
  [hep-th]}}].
%%CITATION = ARXIV:1205.3108;%%

\bibitem{Wolf:2010av}
M.~Wolf,
{\em {A first course on twistors, integrability and gluon scattering
  amplitudes},}
\href{http://dx.doi.org/10.1088/1751-8113/43/39/393001}{J. Phys. A {\bf 43}
  (2010) 393001} [{\tt \href{http://www.arxiv.org/abs/1001.3871}{1001.3871
  [hep-th]}}].

\bibitem{Ward:1990vs}
R.~S.~Ward and R.~O.~Wells,
{\em Twistor geometry and field theory,}
Cambridge University Press, Cambridge, 1990.

\bibitem{Ward:1977ta}
R.~S.~Ward,
{\em {On selfdual gauge fields},}
\href{http://dx.doi.org/10.1016/0375-9601(77)90842-8}{Phys. Lett. A {\bf 61}
  (1977)~81}.
%%CITATION = PHLTA,A61,81;%%

\bibitem{Saemann:2011nb}
C.~Saemann and M.~Wolf,
{\em {On twistors and conformal field theories from six dimensions},}
\href{http://dx.doi.org/10.1063/1.4769410}{J. Math. Phys. {\bf 54} (2013)
  013507} [{\tt \href{http://www.arxiv.org/abs/1111.2539}{1111.2539
  [hep-th]}}].
%%CITATION = ARXIV:1111.2539;%%

\bibitem{Mason:2011nw}
L.~Mason, R.~Reid-Edwards, and A.~Taghavi-Chabert,
{\em {Conformal field theories in six-dimensional twistor space},}
\href{http://dx.doi.org/10.1016/j.geomphys.2012.08.001}{J. Geom. Phys. {\bf 62}
  (2012) 2353} [{\tt \href{http://www.arxiv.org/abs/1111.2585}{1111.2585
  [hep-th]}}].
%%CITATION = ARXIV:1111.2585;%%

\bibitem{Demessie:2014ewa}
G.~A.~Demessie and C.~Saemann,
{\em {Higher Poincar\'e lemma and integrability},}
\href{http://dx.doi.org/10.1063/1.4929537}{J. Math. Phys. {\bf 56} (2015)
  082902} [{\tt \href{http://www.arxiv.org/abs/1406.5342}{1406.5342
  [hep-th]}}].
%%CITATION = ARXIV:1406.5342;%%

\bibitem{Popov:2004rb}
A.~D.~Popov and C.~Saemann,
{\em On supertwistors, the Penrose-Ward transform and $\CN = 4$ super
  Yang-Mills theory,}
\href{http://dx.doi.org/10.4310/ATMP.2005.v9.n6.a2}{Adv. Theor. Math. Phys.
  {\bf 9} (2005) 931} [{\tt
  \href{http://www.arxiv.org/abs/hep-th/0405123}{hep-th/0405123}}].
%%CITATION = HEP-TH/0405123;%%

\bibitem{Saemann:2013pca}
C.~Saemann and M.~Wolf,
{\em {Six-dimensional superconformal field theories from principal 3-bundles
  over twistor space},}
\href{http://dx.doi.org/10.1007/s11005-014-0704-3}{Lett. Math. Phys. {\bf 104}
  (2014) 1147} [{\tt \href{http://www.arxiv.org/abs/1305.4870}{1305.4870
  [hep-th]}}].
%%CITATION = ARXIV:1305.4870;%%

\bibitem{Penrose:1985jw}
R.~Penrose and W.~Rindler,
{\em Spinors and space-time. Vol.\ 1: Two spinor calculus and relativistic
  fields,}
Cambridge University Press, Cambridge, 1984.

\bibitem{Penrose:1986ca}
R.~Penrose and W.~Rindler,
{\em Spinors and space-time. Vol.\ 2: Spinor and twistor methods in space-time
  geometry,}
Cambridge University Press, Cambridge, 1986.

\bibitem{Mason:1991rf}
L.~J.~Mason and N.~M.~J.~Woodhouse,
{\em Integrability, self-duality, and twistor theory,}
Clarendon, Oxford (1996).

\bibitem{Myers:1999ps}
R.~C.~Myers,
{\em Dielectric-branes,}
\href{http://dx.doi.org/10.1088/1126-6708/1999/12/022}{JHEP {\bf 9912} (1999)
  022} [{\tt \href{http://www.arxiv.org/abs/hep-th/9910053}{hep-th/9910053}}].
%%CITATION = HEP-TH/9910053;%%

\bibitem{DeBellis:2010pf}
J.~DeBellis, C.~Saemann, and R.~J.~Szabo,
{\em {Quantized Nambu-Poisson manifolds and $n$-Lie algebras},}
\href{http://dx.doi.org/10.1063/1.3503773}{J. Math. Phys. {\bf 51} (2010)
  122303} [{\tt \href{http://www.arxiv.org/abs/1001.3275}{1001.3275
  [hep-th]}}].
%%CITATION = 1001.3275;%%

\bibitem{Cantrijn:1996aa}
F.~Cantrijn, L.~A.~Ibort, and M.~de~Leon,
{\em Hamiltonian structures on multisymplectic manifolds,}
Rend. Sem. Mat. Univ. Pol. Torino {\bf 54} (1996) 225.

\bibitem{Cantrijn:1999aa}
F.~Cantrijn, A.~Ibort, and M.~de~Leon,
{\em On the geometry of multisymplectic manifolds,}
\href{http://dx.doi.org/10.1017/S1446788700036636}{J. Austr. Math. Soc. {\bf
  66} (1999) 303}.

\bibitem{Baez:2008bu}
J.~C.~Baez, A.~E.~Hoffnung, and C.~L.~Rogers,
{\em {Categorified symplectic geometry and the classical string},}
\href{http://dx.doi.org/10.1007/s00220-009-0951-9}{Commun. Math. Phys. {\bf
  293} (2010) 701} [{\tt \href{http://www.arxiv.org/abs/0808.0246}{0808.0246
  [math-ph]}}].
%%CITATION = ARXIV:0808.0246;%%

\bibitem{Ritter:2015ffa}
P.~Ritter and C.~Saemann,
{\em {Automorphisms of strong homotopy Lie algebras of local observables},}
{\tt \href{http://www.arxiv.org/abs/1507.00972}{1507.00972 [hep-th]}}.
%%CITATION = ARXIV:1507.00972;%%

\bibitem{Ritter:2015ymv}
P.~Ritter and C.~Saemann,
{\em {$L_\infty$-algebra models and higher Chern-Simons theories},}
{\tt \href{http://www.arxiv.org/abs/1511.08201}{1511.08201 [hep-th]}}.
%%CITATION = ARXIV:1511.08201;%%

\bibitem{Bunk:2016rta}
S.~Bunk, C.~Saemann, and R.~J.~Szabo,
{\em {The 2-Hilbert space of a prequantum bundle gerbe},}
{\tt \href{http://www.arxiv.org/abs/1608.08455}{1608.08455 [math-ph]}}.
%%CITATION = ARXIV:1608.08455;%%

\bibitem{Pressley:1988aa}
A.~Pressley and G.~Segal,
{\em Loop groups,} Oxford Mathematical Monographs, Oxford, 1988.

\bibitem{Bergshoeff:2000jn}
E.~Bergshoeff, D.~S.~Berman, J.~P.~{van der Schaar}, and P.~Sundell,
{\em {A noncommutative M-theory five-brane},}
\href{http://dx.doi.org/10.1016/S0550-3213(00)00476-4}{Nucl. Phys. B {\bf 590}
  (2000) 173} [{\tt
  \href{http://www.arxiv.org/abs/hep-th/0005026}{hep-th/0005026}}].
%%CITATION = HEP-TH/0005026;%%

\bibitem{Kawamoto:2000zt}
S.~Kawamoto and N.~Sasakura,
{\em {Open membranes in a constant $C$-field background and noncommutative
  boundary strings},}
\href{http://dx.doi.org/10.1088/1126-6708/2000/07/014}{JHEP {\bf 0007} (2000)
  014} [{\tt \href{http://www.arxiv.org/abs/hep-th/0005123}{hep-th/0005123}}].
%%CITATION = HEP-TH/0005123;%%

\bibitem{Matsuo:0010040}
Y.~Matsuo and Y.~Shibusa,
{\em Volume preserving diffeomorphism and noncommutative branes,}
\href{http://dx.doi.org/10.1088/1126-6708/2001/02/006}{JHEP {\bf 2001} (2001)
  006} [{\tt \href{http://www.arxiv.org/abs/hep-th/0010040}{hep-th/0010040}}].

\bibitem{Saemann:2012ab}
C.~Saemann and R.~J.~Szabo,
{\em {Groupoids, loop spaces and quantization of 2-plectic manifolds},}
\href{http://dx.doi.org/10.1142/S0129055X13300057}{Rev. Math. Phys. {\bf 25}
  (2013) 1330005} [{\tt \href{http://www.arxiv.org/abs/1211.0395}{1211.0395
  [hep-th]}}].
%%CITATION = ARXIV:1211.0395;%%

\bibitem{Waldorf:2007aa}
K.~Waldorf,
{\em Algebraic structures for bundle gerbes and the Wess-Zumino term in
  conformal field theory,} PhD thesis, Universität Hamburg (2007).

\bibitem{Fiorenza:2013kqa}
D.~Fiorenza, C.~L.~Rogers, and U.~Schreiber,
{\em {Higher $\sU(1)$-gerbe connections in geometric prequantization},}
{\tt \href{http://www.arxiv.org/abs/1304.0236}{1304.0236 [math-ph]}}.
%%CITATION = ARXIV:1304.0236;%%

\bibitem{Fiorenza:1304.6292}
D.~Fiorenza, C.~L.~Rogers, and U.~Schreiber,
{\em $L_\infty$-algebras of local observables from higher prequantum bundles,}
\href{http://dx.doi.org/10.4310/HHA.2014.v16.n2.a6}{Homol. Homot. App. {\bf 16}
  (2014) 107} [{\tt \href{http://www.arxiv.org/abs/1304.6292}{1304.6292
  [math-ph]}}].

\bibitem{Saemann:2012ex}
C.~Saemann and R.~J.~Szabo,
{\em {Groupoid quantization of loop spaces},}
PoS {\bf CORFU2011} (2012)~46 [{\tt
  \href{http://www.arxiv.org/abs/1203.5921}{1203.5921 [hep-th]}}].
%%CITATION = ARXIV:1203.5921;%%

\bibitem{Filippov:1985aa}
V.~T.~Filippov,
{\em $n$-Lie algebras,}
\href{http://dx.doi.org/10.1007/BF00969110}{Sib. Mat. Zh. {\bf 26} (1985) 126}.

\bibitem{Cherkis:2008qr}
S.~Cherkis and C.~Saemann,
{\em {Multiple M2-branes and generalized 3-Lie algebras},}
\href{http://dx.doi.org/10.1103/PhysRevD.78.066019}{Phys. Rev. D {\bf 78}
  (2008) 066019} [{\tt \href{http://www.arxiv.org/abs/0807.0808}{0807.0808
  [hep-th]}}].
%%CITATION = ARXIV:0807.0808;%%

\bibitem{deMedeiros:2008zh}
P.~de~Medeiros, J.~M.~Figueroa-O'Farrill, E.~Mendez-Escobar, and P.~Ritter,
{\em {On the Lie-algebraic origin of metric 3-algebras},}
\href{http://dx.doi.org/10.1007/s00220-009-0760-1}{Commun. Math. Phys. {\bf
  290} (2009) 871} [{\tt \href{http://www.arxiv.org/abs/0809.1086}{0809.1086
  [hep-th]}}].
%%CITATION = 0809.1086;%%

\bibitem{Palmer:2012ya}
S.~Palmer and C.~Saemann,
{\em {M-brane models from non-abelian gerbes},}
\href{http://dx.doi.org/10.1007/JHEP07(2012)010}{JHEP {\bf 1207} (2012) 010}
  [{\tt \href{http://www.arxiv.org/abs/1203.5757}{1203.5757 [hep-th]}}].
%%CITATION = ARXIV:1203.5757;%%

\bibitem{Awata:1999dz}
H.~Awata, M.~Li, D.~Minic, and T.~Yoneya,
{\em {On the quantization of Nambu brackets},}
\href{http://dx.doi.org/10.1088/1126-6708/2001/02/013}{JHEP {\bf 02} (2001)
  013} [{\tt \href{http://www.arxiv.org/abs/hep-th/9906248}{hep-th/9906248}}].
%%CITATION = HEP-TH/9906248;%%

\bibitem{Ritter:2013wpa}
P.~Ritter and C.~Saemann,
{\em {Lie 2-algebra models},}
\href{http://dx.doi.org/10.1007/JHEP04(2014)066}{JHEP {\bf 1404} (2014) 066}
  [{\tt \href{http://www.arxiv.org/abs/1308.4892}{1308.4892 [hep-th]}}].
%%CITATION = ARXIV:1308.4892;%%

\end{thebibliography}

\bibliographystyle{latexeu}

\end{document}